\documentclass[superscriptaddress, aps,prd, amsmath,amssymb,showpacs,showkeys, onecolumn]{revtex4-2}
\usepackage[dvips]{graphicx,color}
\usepackage{subfig}
\usepackage{times}
\usepackage{soul}
\usepackage{xcolor}
\usepackage[%
  colorlinks=true,
  urlcolor=blue,
  linkcolor=blue,
  citecolor=blue
]{hyperref}
\usepackage{orcidlink}
\begin{document}

\title{Quasinormal Modes and Greybody Factors of Charged Symmergent Black Hole }

\author{Dhruba Jyoti Gogoi \orcidlink{0000-0002-4776-8506}}
\email{moloydhruba@yahoo.in}
\affiliation{Department of Physics, Moran College, Moranhat, Charaideo 785670, Assam, India.}
\affiliation{Theoretical Physics Division, Centre for Atmospheric Studies, Dibrugarh University, Dibrugarh
786004, Assam, India.}

\author{Beyhan Puli\c{c}e
\orcidlink{0000-0002-5277-6906}
}
\email{beyhan.pulice@sabanciuniv.edu}
\affiliation{Faculty of Engineering and Natural Sciences, Sabanc{\i} University, 34956 Tuzla, \.{I}stanbul, Turkiye.}
\affiliation{Astrophysics Research Center (ARCO), the Open University of Israel, Raanana 4353701, Israel.}

\author{Ali \"Ovg\"un \orcidlink{0000-0002-9889-342X}}
\email{ali.ovgun@emu.edu.tr}
\affiliation{Physics Department, Eastern Mediterranean University, Famagusta, 99628 
North Cyprus via Mersin 10, Turkiye.}

\begin{abstract}
In this paper, we investigate quasinormal modes (QNMs) and greybody factors within the framework of Symmergent gravity, an emergent gravity model with an $R + R^2$ curvature sector. Building on our previous work on static spherically-symmetric solutions [Class.Quant.Grav. 40 (2023) 19, 195003], we explore the effects of the key parameters, including the quadratic curvature coupling parameter $c_{\rm O}$ and the vacuum energy parameter $\alpha$. For both scalar and electromagnetic perturbations, an increase in $\alpha$ leads to a nearly linear rise in both oscillation frequencies and damping rates. The other parameter $c_{\rm O}$ affects the QNMs spectrum nonlinearly. Additionally, the charge $Q$ of the black hole introduces nonlinear behavior, where higher charges amplify the black hole's electromagnetic field, resulting in increased oscillation frequencies and faster stabilization. These findings enhance our understanding of charged black hole stability and gravitational wave astrophysics. 
Further, the analysis of greybody factors reveals that increasing $\alpha$, $c_{\rm O}$, and $Q$ reduces the absorption of radiation, with electromagnetic perturbations reaching maximum absorption at slightly lower frequencies compared to scalar perturbations.  

\end{abstract}
\date{\today}

\keywords{charged Symmergent black hole; quasinormal modes; ring-down gravitational waves; greybody factors.}
\pacs{04.40.-b, 95.30.Sf, 98.62.Sb}

\maketitle

%\tableofcontents
\section{Introduction}
The Standard Model (SM) describes the strong, weak, and electromagnetic interactions as a renormalizable quantum field theory (QFT) \cite{wein0}, but does not account for gravity. It is unique to the flat Minkowski spacetime because of difficulties in quantizing the curved metric \cite{quant-gravity} and in transferring QFTs into curved spacetime \cite{weinx2,weinx3}. In contrast to full QFTs like the SM, flat spacetime effective QFTs are expected to exhibit a closer connection to curved spacetime. This is because they are nearly classical field theories obtained by integrating out high-frequency quantum fluctuations \cite{wein0,wein1}, in the framework of both Wilsonian and one-particle-irreducible effective actions \cite{polchinski, burgess}.

The SM is endowed with a physical UV cutoff (rather than a regulator) when viewed as an effective QFT \cite{wein1}. This UV cutoff denoted by $\Lambda_\wp$, represents the scale at which a UV completion becomes relevant. As revealed by the null LHC searches \cite{lhc}, $\Lambda_\wp$ could be at any scale above a TeV. In general, in a Lorentz conserving regime, the loop momentum $\ell^\mu$ lies in the interval $-\Lambda_\wp^2 \leq \eta_{\mu\nu} \ell^\mu \ell^\nu \leq \Lambda_\wp^2$, and it explicitly breaks the Poincar\'{e} (translation) symmetry. In the presence of the UV cutoff $\Lambda_\wp^2$, scalar mass-squareds obtain corrections proportional to  $\Lambda_\wp^2$, and $\Lambda_\wp^2$ and $\Lambda_\wp^4$ terms appear in the loop-corrected vacuum energy \cite{wein1,veltman}. Beyond these UV sensitivities \cite{wein1}, gauge bosons also acquire mass-squared terms proportional to $\Lambda_\wp^2$, and which leads to a breaking of gauge symmetries \cite{gauge-break1,gauge-break2}. In order to alleviate these unnatural UV sensitivities, the most sensible line of action would be to restore the gauge symmetries by means of the Higgs mechanism \cite{Higgs1, Higgs2, Higgs3}. A hindrance to this proposal is the stark difference between the intermediate vector boson mass (Poincar\'{e}-conserving) \cite{wein2}  and the loop-induced gauge boson mass (Poincar\'{e}-breaking) \cite{dad2023}. Indeed, in the former, the vector boson mass is promoted to a scalar field, which leads to the usual Higgs mechanism \cite{Higgs1, Higgs2, Higgs3}. In the latter, however, it is necessary to first find a Poincar\'{e}-breaking Higgs field. In this regard, it turns out that the affine curvature, which is independent of the metric tensor and its connection, is the sought-after Higgs field in that it promotes the UV cutoff  $\Lambda_\wp$  \cite{affine1,affine2} and condenses to the usual metrical curvature in the minimum of the metric-affine action \cite{affine3,affine4}. The defusion of the unnatural UV sensitivity, the restoration of gauge symmetries, the emergence of gravity, and the appearance of new particles without the necessity to couple directly to SM particles are the salient outcomes of this condensation. This whole mechanism is referred to as gauge symmetry-restoring emergent gravity or {\it symmergent gravity} in brief. Symmergent gravity has been constructed and applied through various stages  \cite{dad2023,dad2021,dad2019,dad2024}, starting with the nascent idea of gauge symmetry restoration by means of curvature \cite{dad2016}. Some applications of Symmergent gravity can be found in \cite{irfanaliben,Symmergent-bh2,Pulice:2024wjg,Pulice:2023dqw,dark-sector,Symmergent-bh2,reggiealiben,cqgirfan,bh4,inf1}.

Black holes are not isolated cosmic bodies; rather, they are dynamic entities that actively interact with their surroundings, profoundly influencing the fabric of spacetime. When a black hole is perturbed, spacetime itself ripples, creating waves that eventually die down as the system settles back to equilibrium. Understanding how these systems respond to small disturbances is a fundamental aspect of physics, providing deep insights into the nature of spacetime and gravity. The study of black hole perturbations began with pioneering work by \cite{Regge:1957td}, and was significantly advanced by others \cite{Vishveshwara:1970cc,Zerilli:1970se,Zerilli:1970wzz,Zerilli:1974ai,Moncrief:1975sb,Teukolsky:1972my}, leading to Chandrasekhar’s seminal contributions \cite{Chandrasekhar:1985kt,Chandrasekharbook}. At the heart of this field are quasinormal modes (QNMs), which are complex frequencies characterizing the oscillatory behavior of a black hole as it returns to a stable state after being disturbed. These modes are unique signatures that depend on the black hole’s intrinsic properties, such as its mass, charge, and spin, as well as the nature of the perturbation—whether it is scalar, electromagnetic, or gravitational. The independence of QNMs from initial conditions makes them particularly valuable in probing the characteristics of black holes, offering a direct link between theoretical predictions and observable phenomena \cite{Konoplya:2011qq,Berti:2009kk,Berti:2005ys,Ferrari:2007dd,Daghigh:2008jz,Daghigh:2020jyk,Daghigh:2020mog,Daghigh:2006xg,Daghigh:2022uws,Fernando:2016ftj,Fernando:2015kaa,Fernando:2012yw,Fernando:2008hb,Chabab:2016cem,Chabab:2017knz,Lepe:2004kv,Gonzalez:2017shu,Khodadi:2018scn,Gogoi:2022wyv,Gogoi:2023kjt,Gogoi:2021cbp,Gogoi:2021dkr,Oliveira:2021abg,Kuang:2017cgt,Kac:1966xd,Andersson:1992scr,Andersson:1994rm,Andersson:1995vi,Andersson:1996xw,Maggio:2019zyv,Konoplya:2022zym,Konoplya:2018yrp,Cardoso:2016rao,Cardoso:2016oxy,Visinelli:2017bny,Vagnozzi:2020gtf,Casalino:2018wnc,Casalino:2018tcd,Cardoso:2017cqb,Vagnozzi:2023lwo,Benetti:2021uea,Yang:2019vni,Pantig:2022gih,Lambiase:2023hng,Yang:2022ifo,Okyay2022,Daghigh:2011ty,Zhidenko:2003wq,Zhidenko:2005mv,Gonzalez:2022ote,Rincon:2018ktz}.

The study of QNMs is especially critical in the context of the ringdown phase of binary black hole mergers. During this phase, the newly formed black hole emits gravitational waves as it stabilizes, with the QNMs encoding key information about the black hole’s parameters. As gravitational wave astronomy continues to grow, understanding QNMs has become increasingly important for interpreting the data from detectors like LIGO and Virgo \cite{LIGOScientific:2016aoc,LIGOScientific:2016sjg}.

Complementing the study of QNMs is the analysis of the greybody factor, which describes the modification of the spectrum of radiation as it escapes the black hole’s gravitational well. The greybody factor plays a crucial role in understanding the energy distribution of radiation emitted by black holes, influencing the signals detected by gravitational wave observatories. Together, QNMs and the greybody factor provide a comprehensive picture of black hole dynamics and their interactions with the surrounding environment.

In light of this, our study focuses on computing QNMs and greybody factors for charged symmergent black holes (CSBHs). By examining these scenarios, we aim to uncover novel insights into the behavior of black holes in complex electromagnetic environments, contributing to a deeper understanding of these enigmatic objects in the universe.

Our work is organized as follows: after the introduction, we briefly review the CSBHs in Section II. In Sections III and IV, we investigate massless scalar and massless vector perturbations around the black hole. In Section V, we study greybody factor using the WKB approach and rigorous bounds. Finally, we summarize and conclude our paper in Section VI.

\section{Brief Review of Charged Symmergent Black Holes} \label{sec3}

Symmergent gravity is a specific form of emergent gravity that incorporates both $ R $ and $ R^2 $ curvature terms along with a non-zero cosmological constant. As a specialized case within the broader class of $ f(R) $ gravity theories, it was first introduced in \cite{dad2019,dad2016}, with further refinements and enhancements presented in \cite{dad2021}. This model has been applied across various contexts \cite{irfanaliben,Symmergent-bh2,Pulice:2023dqw}, while recent field-theoretic and string-theoretic developments have been detailed in \cite{dad2024}. In fact, symmergent gravity is defined by the action 
\begin{eqnarray}
\label{curvature-sector}
S[g]= \int d^4x \sqrt{-g}\left\{
\frac{R}{16\pi G} - \frac{c_{\rm O}}{16} R^2 - V_{\rm O}  + {\mathcal{L}}_{matter}\right\},
\end{eqnarray}
in which $R$ is the Ricci curvature scalar and ${\mathcal{L}}_{matter}$ represents the Lagrangian for both standard model matter fields (quarks, leptons, gauge bosons, and the Higgs) and additional fields required to generate the inverse Newton's constant at one loop in the flat spacetime effective action. The relationship for the inverse of Newton's constant is given by
\begin{eqnarray}
\label{params-0}
\frac{1}{G} = \frac{{\rm str}\left[{\mathcal{M}}^2\right]}{8 \pi},
\end{eqnarray}
which is set by the mass-squared matrix ${\mathcal{M}}^2$ of the entirety of matter fields such that its supertrace ${\rm str}[{\mathcal{M}}^2] = \sum_i (-1)^{2s_i+1} (2 s_i +1) {\rm tr}[{\mathcal{M}}^2]_{s_i}$ runs over all fields $\psi_i$ with spin $s_i$. At the one-loop level, the quadratic curvature coefficient in Eq.~\eqref{curvature-sector} is given by the number difference between the total bosonic degrees of freedom $n_B$ and the fermionic degrees of freedom $n_F$ in the underlying QFT \cite{dad2023}, namely 
\begin{eqnarray}
\label{params}
 c_{\rm O} = \frac{n_\text{B} - n_\text{F}}{128 \pi^2}\,.
\end{eqnarray}
Similarly, again at the one-loop level, the vacuum energy in Eq.~\eqref{curvature-sector} is given by
\begin{align}
\label{params-vacuum}
V_{\rm O} = \frac{{\rm str}\left[{\mathcal{M}}^4\right]}{64 \pi^2}.
\end{align}
The loop-induced parameters $ n_\text{B} $ and $ n_\text{F} $ play a crucial role in determining key aspects of the Symmergent gravity. These degrees of freedom are not limited to the known particles of the standard model; they also encompass entirely new particles, both massive and massless, that may not interact with the SM particles non-gravitationally. This broader inclusion of particle species highlights the flexibility of Symmergent gravity, allowing it to incorporate new physics beyond the SM without requiring couplings to the known particles. Consequently, the contributions from these additional particles affect the gravitational sector primarily through their impact on loop corrections, influencing the effective gravitational dynamics.

In the context of general $n_\text{B}$ and $n_\text{F}$, the Symmergent gravity action (\ref{curvature-sector}) may be expressed in the $f(R)$ gravity form as shown below \cite{Pulice:2023dqw}: 
\begin{eqnarray}
S[g]=\frac{1}{16 \pi G} \int d^4 x \sqrt{-g} \left(f(R) - 2 \Lambda-\frac{1}{2} \hat{F}_{\mu \nu} \hat{F}^{\mu \nu} \right)
\label{fr-action1}
\end{eqnarray}
where $f(R)= R+ \beta R^2 $
with the quadratic curvature coefficient 
$\beta = - \pi G c_{\rm O}\,$
and the cosmological constant 
$\Lambda=8 \pi G V_{\rm O}.$
A detailed analysis of the vacuum energy $V_{\rm O}$ starting from its definition in Eq. (\ref{params-vacuum}), will be presented in the subsequent part of the paper. 

In this paper, we focus solely on the electromagnetic field from the matter sector in the action (\ref{fr-action1}). The electromagnetic field tensor is given by
\begin{align}
\hat{F}_{\mu \nu} = \partial_{\mu} \hat{A}_{\nu} -\partial_{\nu} \hat{A}_{\mu} 
\end{align}
where the dimensionless electromagnetic potential is defined as $\hat{A}_\mu = A_\mu/\sqrt{8\pi G}$.

In this section, we derive the charged black hole solution for the combined Symmergent gravity and Maxwell system from the action in (\ref{fr-action1}). The gravitational field equations are expressed as follows,
\begin{align}
\label{Einstein-eqns}
E_{\mu\nu}\equiv R_{\mu \nu} F(R)-\frac{1}{2} g_{\mu \nu} f(R) + g_{\mu \nu} \Lambda + (g_{\mu \nu} \square -\nabla_{\mu} \nabla_{\nu}) F(R) - {\hat T}_{\mu \nu} = 0
\end{align} 
and are coupled with the Maxwell field equations:
\begin{align}
\label{Maxwell-eqn}
\partial_\mu (\sqrt{-g} \hat{F}^{\mu \nu}) &= 0
\end{align}
where $F(R)\equiv d f(R)/dR$ and ${\hat T}_{\mu\nu}$, the energy-momentum tensor of the dimensionless Maxwell field in (\ref{Einstein-eqns}), is given by:
\begin{align}
{\hat T}_{\mu \nu}=g^{\alpha \beta} \hat{F}_{\alpha \mu}  \hat{F}_{\beta \nu}-\frac{1}{4}  g^{\gamma \alpha} g^{\rho \beta} \hat{F}_{\alpha \beta} \hat{F}_{\gamma \rho}.    
\end{align}

We now seek a static, spherically symmetric solution for the combined Symmergent gravity and Maxwell system. For this, we propose the following metric:  
\begin{align}
d s^{2} = -h(r) d t^{2}+\frac{1}{h(r)} d r^{2} + r^{2} (d \theta^{2} +  \sin ^{2} \theta d \phi^{2})
\label{metric-fR}, 
\end{align}
where $ h(r) $ is the single metric potential. The corresponding electromagnetic scalar potential is $ \hat{A}_0 = \hat{q}(r) $, with the vector potential vanishing ($ \hat{A}_i = 0 $, for $ i = 1, 2, 3 $). The electrostatic potential is given by: 
$\hat{q}(r) = \frac{Q}{r}$
where a homogeneous term has been discarded, ensuring the scalar potential follows a pure Coulomb form. For a QFT characterized by a scale $ M_0 $ but lacking detailed knowledge of the mass spectrum, realistic scenarios can be modeled by introducing the parametrization:  
\begin{eqnarray}
\label{analyze-VO-2}
V_{\rm O} &=& \frac{1-{\alpha}}{(8\pi G)^2 c_{\rm O}},
\end{eqnarray}
where the parameter $ \alpha $ reflects deviations in the boson and fermion masses from the characteristic scale $ M_0 $. In general, $ \alpha > 1 $ ($ \alpha < 1 $) indicates fermion (boson) dominance in terms of the trace of $ (\text{masses})^4 $, with $ \alpha > 1 $ corresponding to an AdS spacetime and $ \alpha < 1 $ corresponding to a dS spacetime.

With the vacuum energy given by Eq. (\ref{analyze-VO-2}), the metric potential $h(r)$  takes the form \cite{Pulice:2023dqw}:
\begin{eqnarray}
h(r) = 1 - \frac{2MG}{r} + 
\frac{Q^2}{2 \alpha r^2} -\frac{(1-\alpha)}{24 \pi G c_{\rm O}} r^2 \, .
\label{metric}
\end{eqnarray} 
It results in the standard Reissner-Nordström-AdS/dS black hole when $\hat{Q}^2 =\frac{Q^2}{2\alpha}$ and $\hat{\Lambda}=\frac{(1-\alpha)}{8 \pi G c_{\rm O}}$. The metric (\ref{metric-fR}) with the potential (\ref{metric}) is called as CSBH \cite{Pulice:2023dqw}.

The predictions of the Einstein field equations have been tested extensively in a broad range of observationally accessible scenarios \cite{W:rev:14}, including but not limited to inflationary evolution \cite{inf1}, the formation of neutron stars and black holes \cite{irfanaliben,Symmergent-bh2,reggiealiben,bh4,bh5,cqgirfan,Pulice:2023dqw,Pulice:2024wjg}, and the emission of gravitational waves \cite{gw1,gw2}. Given the increase in the number and variety of astrophysical observations of ultracompact objects (inlcuding recent EHT \cite{EventHorizonTelescope:2022xnr} and JWST \cite{uncover} observations), black holes manifest themselves as viable testbeds. This motivates our study of black hole spacetimes depending on the parameters $c_{\rm O}$ and $\alpha$.

\section{Perturbations and Quasinormal modes}\label{sec3} 
In this section, we focus on the analysis of massless scalar and vector perturbations within the framework of the black hole metric defined in equation \eqref{metric}. For the sake of simplicity, we assume that the test field, whether a scalar field or a vector field, exerts a negligible influence on the black hole spacetime, meaning that the back reaction of the field on the spacetime geometry is insignificant \cite{lopez2020,Chandrasekhar:1985kt}. This assumption allows us to treat the perturbations as propagating on a fixed background metric without needing to consider the complex, coupled dynamics that would arise if the back reaction were significant.

To investigate the QNMs associated with these perturbations, we derive Schr\"odinger-like wave equations tailored to each case by imposing the relevant conservation relations on the background spacetime. For scalar fields, this process results in equations of the Klein-Gordon type, which describe how the scalar perturbations evolve in the curved spacetime. For electromagnetic (vector) fields, the corresponding wave equations are derived from Maxwell's equations, governing the dynamics of electromagnetic perturbations. These equations encapsulate the key physical properties of the perturbations, such as their frequencies and damping rates, which are essential for determining the QNMs.

The QNMs are calculated using the Pad\'e-averaged 6th-order WKB approximation method, a semi-analytical technique that is particularly effective for evaluating the frequencies and decay rates of perturbations in black hole spacetimes. The WKB method is well-suited for handling wave equations with a potential barrier, as is typical in the context of black hole perturbations. By employing this method, we can obtain accurate estimates of the QNMs, which provide insights into the stability and dynamical response of the black hole to perturbations.

Our analysis specifically focuses on axial perturbations. To describe these perturbations, we express the perturbed metric as shown in the following way \cite{lopez2020} 
\begin{equation}  \label{pert_metric}
ds^2 = -g_{tt} dt^2 +
r^2 d\theta^2  + g_{rr} dr^2 + r^2 \sin^2\!\theta\, \left (d\phi - p_1(t,r,\theta)\,
dt - p_2(t,r,\theta)\, dr - p_3(t,r,\theta)\, d\theta \right )^2 \, ,
\end{equation}
where the functions $p_1(t,r,\theta)$, $p_2(t,r,\theta)$, and $p_3(t,r,\theta)$ characterize the perturbations in the black hole spacetime. These functions represent deviations from the static and spherically symmetric background metric, with $g_{tt} = h(r)$ and $g_{rr} = 1/h(r)$ serving as the zeroth-order terms in the perturbative expansion. These zeroth-order terms correspond to the unperturbed black hole metric, while the functions $p_i$ encode the effects of the axial perturbations.

Through this approach, we systematically study the impact of massless scalar and vector perturbations on the black hole spacetime, providing a deeper understanding of the behavior and characteristics of QNMs in this context. This analysis is crucial for exploring the stability of black holes and the nature of gravitational and electromagnetic waves in curved spacetime, offering valuable insights into the fundamental properties of black hole physics.

\subsection{Massless Scalar Perturbation}
\label{massless-scalar}

We begin by considering a massless scalar field $\Phi$ in the vicinity of a CSBH, governed by the equation $\square \Phi = 0$, which describes the motion of the scalar field assuming negligible backreaction. The corresponding Klein-Gordon equation in the CSBH spacetime takes the form
\begin{equation}
\dfrac{1}{r^2 \sqrt{g_{tt} g_{rr}^{-1}}} \left[r^2 \sqrt{g_{tt} g_{rr}^{-1}} \Phi_{,r} \right]_{,r} + \dfrac{g_{tt}}{r^2\sin\theta}\left(\sin\theta \Phi_{,\theta}\right)_{,\theta}+ \dfrac{g_{tt}}{r^2 \sin^2\theta} (\Phi)_{,\phi\phi} - (\Phi)_{,tt} = 0.    
\end{equation}

The angular structure of this equation suggests that the appropriate basis functions for the polar orientation are the associated Legendre polynomials $P_l^m(\cos\theta)$, which satisfy the relation
\begin{equation}
\dfrac{1}{\sin\theta} \dfrac{d}{d\theta}\left(\sin\theta \frac{d P_l^m(\cos\theta)}{d\theta}\right) - \frac{m^2}{\sin^2\theta}\Phi = - l(l+1) P_l^m(\cos\theta).    
\end{equation}

After separating the $\phi$-dependent part of the equation using $\partial^2_\phi \Phi = -m^2 \Phi$, we express the scalar perturbation as a multipole expansion:
\begin{equation}
\Phi(t,r,\theta,\phi) = \dfrac{1}{r} \sum_{l,m} \sqrt{\dfrac{(2l+1)}{4\pi} \dfrac{(l-m)!}{(l+m)!}} \psi_{s}(t,r) e^{im\phi} P_l^m (\cos\theta),    
\end{equation}
where $l$ and $m$ represent the polar and azimuthal indices, respectively \cite{sommerfeld,Chandrasekhar:1985kt}. The function {$\psi_{s}(t,r)$}  
is the radial time-dependent wave function and the index ``s" stands for scalar.

Substituting this expansion into the Klein-Gordon equation leads to the stationary Schrödinger-like equation for the radial part:
\begin{equation}
    \partial^2_{r_*} \psi(r_*)_{sl} + \omega^2 \psi(r_*)_{s} = { V_s(r)} \psi(r_*)_{s},
\end{equation}
where $r_*$ is the tortoise coordinate. 

The term $V_s(r)$ represents the effective potential, given by
\begin{equation}
   V_s(r) = |g_{tt}| \left( \dfrac{l(l+1)}{r^2} + \dfrac{1}{r \sqrt{|g_{tt}| g_{rr}}} \dfrac{d}{dr}\sqrt{|g_{tt}| g_{rr}^{-1}} \right) = h(r) \left( \dfrac{l(l+1)}{r^2} + \dfrac{1}{r} \dfrac{d}{dr}h(r) \right). 
\end{equation}
Here, $l$ corresponds to the multipole moment of the black hole's QNMs, and $\omega$ represents the frequency of these modes, which is determined from {$\partial_t^2 \psi_{s}(t,r_* ) = -\omega^2 \psi_{s}(t,r_* )$}. The wavefunction $\psi(r_* )_s $ thus describes behaviour of the scalar QNMs of the CSBH \cite{Chandrasekhar:1985kt, lopez2020}. 

This formalism shows how scalar field perturbations evolve in the spacetime around CSBHs, with the effective potential and tortoise coordinate playing key roles in shaping the dynamics of the QNMs. The diagonal and static nature of the CSBH metric facilitates the straightforward definition of the QNM frequencies, which are essential for understanding the behavior of the scalar perturbations in these exotic spacetimes.

\subsection{Massless Vector or Electromagnetic Perturbation}
\label{massless-vector}
Following the analysis of scalar perturbations, we now turn to the study of massless vector perturbations, specifically focusing on electromagnetic perturbations. The tetrad formalism proves to be a useful approach in analyzing these perturbations, as it allows the projection of the curved metric $ g_{\mu\nu} $ onto the flat metric $ \eta_{\bar{\mu} \bar{\nu}} $ using the vierbein $ e_\mu^{\bar{\mu}} $. In this formalism, the relationship between the curved and flat metrics is given by $ g_{\mu\nu} = e_\mu^{\bar{\mu}} \eta_{\bar{\mu} \bar{\nu}} e^{\bar{\nu}}_\nu $, where the indices $\mu, \nu, \dots$ refer to the curved spacetime, and $\bar{\mu}, \bar{\nu}, \dots$ refer to the flat spacetime. The vierbeins satisfy several key relations $e^{\bar \mu}_\mu e^\mu_{\bar \nu} = \delta^{\bar \mu}_{\bar \nu}$, $e^{\bar \mu}_\mu e^\nu_{\bar \mu} = \delta^{\nu}_{\mu}$ and $e^{\bar \mu}_\mu = g_{\mu\nu} \eta^{\bar\mu \bar\nu} e^\nu_{\bar\nu}$ enabling us to express a given vector $ S_\mu $ and a tensor $ P_{\mu\nu} $ in terms of their flat-spacetime counterparts.

The electromagnetic field strength tensor $ F_{\bar{\mu} \bar{\nu}} $ is governed by the Jacobi identity $ \partial_{[\bar{\mu}} F_{\bar{\nu} \bar{\rho}]} = 0 $, leading to a set of constraints on the vector perturbations. %These constraints, represented by equations \eqref{em1} and \eqref{em2}, describe the behavior of the electromagnetic field components. Additionally, a third relation, equation \eqref{em3}, follows from electric charge conservation and provides further insight into the system's dynamics.
Using these equations, we derive a two-derivative motion equation for the vector perturbations, expressed as
\begin{equation}
    \left[ \sqrt{g_{tt} g_{rr}^{-1}} \left( r \sqrt{g_{tt}}\, \mathcal{F} \right)_{,r} \right]_{,r} + \dfrac{g_{tt} \sqrt{g_{rr}}}{r} \left( \dfrac{\mathcal{F}_{,\theta}}{\sin\theta} \right)_{,\theta} \sin\theta - r \sqrt{g_{rr}}\, \mathcal{F}_{,tt} = 0,
\end{equation}
where $\mathcal{F} = F_{\bar{t} \bar{\phi}} \sin\theta$. Unlike scalar perturbations, the angular part of this equation suggests that the appropriate basis functions are the Gegenbauer polynomials $ P_l(\cos\theta | -1) $, which satisfy the differential equation
\begin{equation}
    \sin\theta \dfrac{d}{d\theta}\left(\dfrac{1}{\sin\theta} \frac{d P_l(\cos\theta | -1)}{d\theta}\right) = - l(l-1) P_l(\cos\theta | -1).
\end{equation}

The vector wavefunction $\mathcal{F}(t, r, \theta, \phi)$ can then be expanded in terms of these polynomials as
\begin{equation}
    \mathcal{F}(t,r,\theta, \phi) = \sum_l {\psi_{em}(t,r)} \dfrac{dP_l(\cos\theta | -1)}{\sin\theta d\theta},
\end{equation}

leading to the equation for the radial part of the vector perturbations:
\begin{equation}
    \left[ \sqrt{g_{tt} g_{rr}^{-1}} \left( r \sqrt{g_{tt}}\, \psi_{em} \right)_{,r} \right]_{,r} + \omega^2 r \sqrt{g_{rr}}\, \psi_{em} - \dfrac{1}{r} g_{tt} \sqrt{g_{rr}}\, l(l+1)\, \psi_{em} = 0.
\end{equation}
Here $\psi_{em}$ is the wave function associated with the electromagnetic perturbation.
This equation mirrors the form of the scalar perturbation equation, with the key difference being the absence of a derivative term in the effective potential for vector perturbations.

By defining the tortoise coordinate $ r_* $ as in the scalar case, the vector perturbations obey the stationary Schrödinger equation
\begin{equation}
    \partial^2_{r_*} \psi_{em}(r_*) + \omega^2 \psi_{em}(r_*) = V_{em}(r) \psi_{em}(r_*),
\end{equation}
where the effective potential for electromagnetic perturbations is
\begin{equation}
    V_{em}(r) = g_{tt} \dfrac{l(l+1)}{r^2} = h(r) \dfrac{l(l+1)}{r^2}.
\end{equation}

This potential differs from that of scalar perturbations, as it lacks the derivative term associated with the metric function $ h(r) $.

While these vector perturbations can be identified with the electromagnetic field, the broader symmergent gravity framework predicts the existence of additional massless particles, which may not interact with known particles through non-gravitational forces. These could include massless scalar fields or massless (dark) photons, which could significantly influence gravitational wave emissions. The presence of such particles in the symmergent particle spectrum could lead to observable deviations in the dynamics of gravitational waves and other astrophysical phenomena \cite{dad2019,dad2023,dad2021}.

\subsection{Behaviour of the Perturbation Potentials}

Here, we briefly explore the characteristics of the perturbation potentials associated with the black hole under consideration. The perturbation potential plays a crucial role in determining the QNMs of the black hole, as the nature of the potential directly influences the frequencies and damping rates of these oscillatory modes. By analyzing the behavior of the potential, we can gain a preliminary understanding of the QNMs and their dependence on various parameters.

\begin{figure}[!h]
\centerline{
   \includegraphics[scale = 0.8]{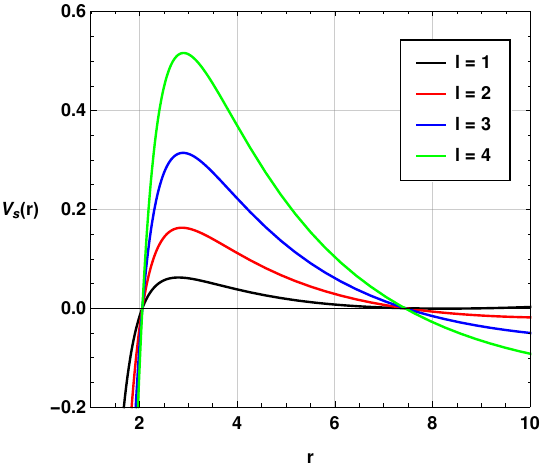}\hspace{0.5cm}
   \includegraphics[scale = 0.8]{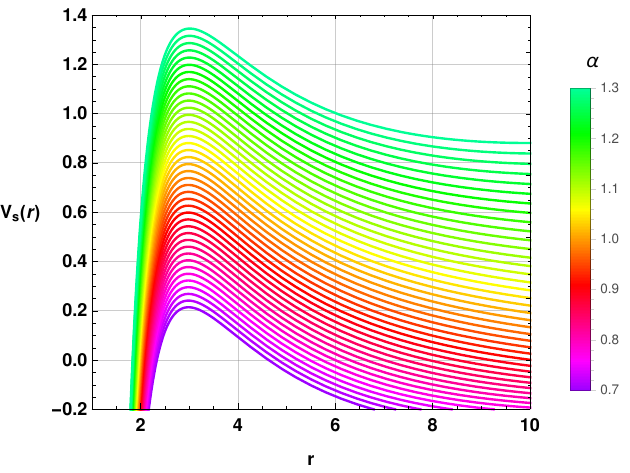}} \vspace{-0.2cm}
\caption{Variation of the scalar potential $V_s(r)$ of CSBH with the radial distance $r$ for different values of the multipole moment $l$ with $M=1$, $G = 1$, $\alpha = 0.7$, $Q = 0.4$ and $c_{\rm O} = 0.3$ (left panel), and for different values of $\alpha$ with $l=4$, $M = 1$, $Q = 0.1$ and $c_{\rm O} = 0.15$ (right panel). }
\label{fig_Vs_01}
\end{figure}

\begin{figure}[!h]
\centerline{
   \includegraphics[scale = 0.8]{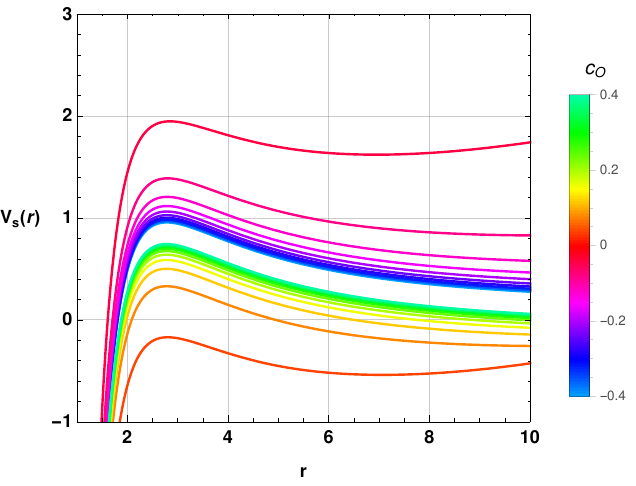}
   \hspace{0.5cm}
   \includegraphics[scale = 0.8]{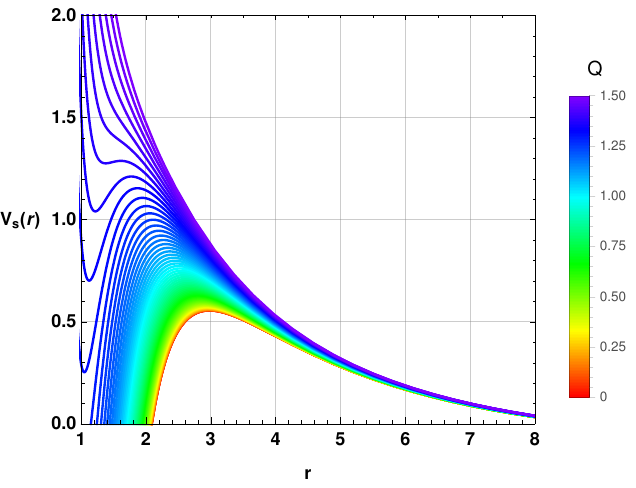}}\vspace{-0.2cm}
\caption{Variation of the scalar potential $V_s(r)$ of CSBH with the radial distance $r$ with $M=1$, $G = 1$, $\alpha = 0.85$ and $l=4$. On the left panel $Q = 0.7$. On the right panel $c_{\rm O} = 0.2$.  }
\label{fig_Vs_02}
\end{figure}

\begin{figure}[!h]
\centerline{
   \includegraphics[scale = 0.8]{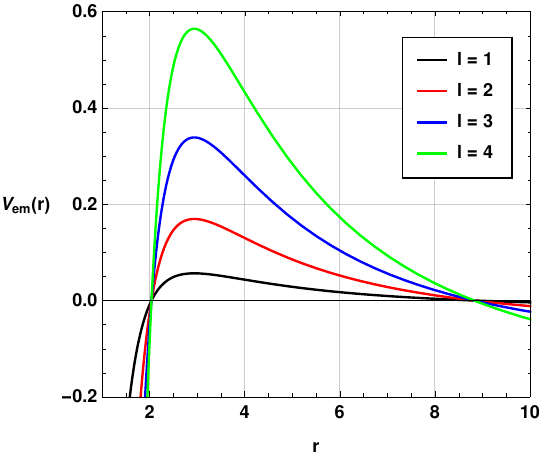}\hspace{0.5cm}
   \includegraphics[scale = 0.8]{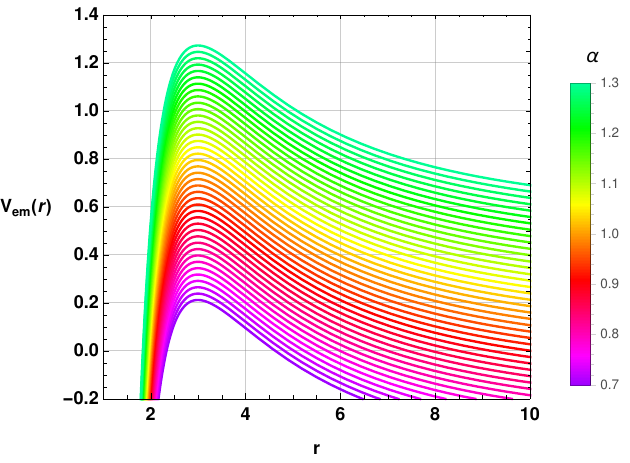}} \vspace{-0.2cm}
\caption{Variation of electromagnetic potential $V_{em}(r)$ of CSBH with the radial distance $r$ for different values of the multipole moment $l$ with $M=1$, $G = 1$, $\alpha = 0.7$, $Q = 0.35$ and $c_{\rm O} = 0.4$ (left panel), and for different values of $\alpha$ with $M = 1$, $l=4$, $Q = 0.1$ and $c_{\rm O} = 0.15$ (right panel). }
\label{fig_Vem_01}
\end{figure}

\begin{figure}[!h]
\centerline{
   \includegraphics[scale = 0.8]{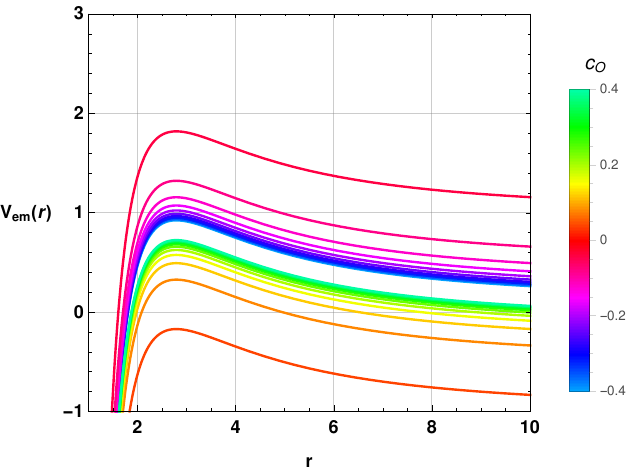}
   \hspace{0.5cm}
   \includegraphics[scale = 0.8]{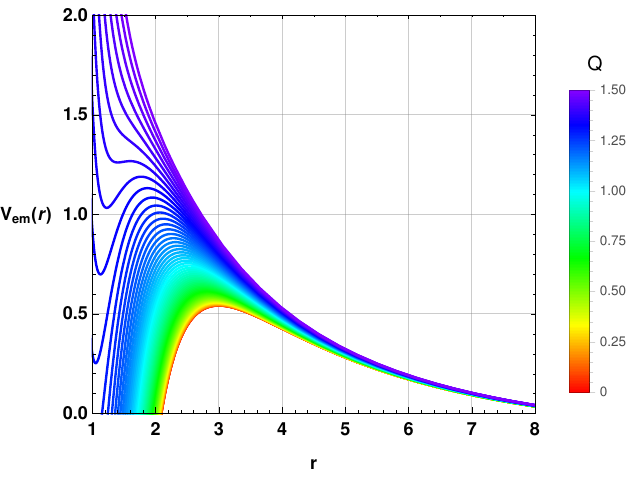}}\vspace{-0.2cm}
\caption{Variation of electromagnetic potential $V_{em}(r)$ of CSBH with the radial distance $r$ with $M = 1$, $l=4$ and $\alpha = 0.85$. On the left panel $Q = 0.7$. On the right panel $c_{\rm O} = 0.2$. }
\label{fig_Vem_02}
\end{figure}

The first panel of Fig. \ref{fig_Vs_01} illustrates the scalar potential for different values of the multipole moment $l$. As expected, the peak value of the potential increases with an increase in $l$. This trend is consistent with the understanding that higher multipole moments correspond to more complex angular dependencies in the perturbation, leading to a stronger potential barrier. The rise in the potential peak with increasing $l$ indicates that the black hole is more resistant to perturbations with higher angular momentum, which will likely result in QNMs with higher frequencies and shorter lifetimes. This is because the potential barrier effectively traps the perturbation closer to the black hole, leading to faster decay and higher oscillation frequencies.

In the second panel of Fig. \ref{fig_Vs_01}, we examine the variation of the scalar potential with respect to the model parameter $\alpha$. The potential curve rises significantly as $\alpha$ increases, indicating that this parameter has a substantial impact on the behavior of the potential. The parameter $\alpha$ likely represents a characteristic of the modified gravity model or an additional field interacting with the black hole spacetime. The increase in the potential with $\alpha$ suggests that higher values of $\alpha$ make the black hole more resistant to scalar perturbations, potentially leading to higher QNM frequencies and faster damping rates. This observation underscores the sensitivity of the QNMs to the underlying model parameters, highlighting the importance of understanding how these parameters influence the perturbation dynamics.

Moving to Fig. \ref{fig_Vs_02}, the first panel displays the variation of the scalar potential concerning the parameter $c_{\rm O}$. The potential curve shows a rapid increase for very small positive values of $c_{\rm O}$, while for negative values close to zero, the potential decreases drastically. As $c_{\rm O}$ becomes more positive or more negative, the potential tends to stabilize towards a central value. This behavior indicates that $c_{\rm O}$ exerts a complex influence on the potential, possibly related to the asymptotic behavior of the spacetime or the nature of the coupling between the black hole and the scalar field. The rapid changes in the potential for small values of $c_{\rm O}$ suggest that even slight variations in this parameter can significantly alter the QNMs, making it a critical factor in the stability analysis of the black hole.

The second panel of Fig. \ref{fig_Vs_02} shows the variations of the potential with respect to the charge $Q$ of the black hole. The black hole charge has a pronounced effect on the potential behavior. As $Q$ increases, the peak value of the potential rises significantly, and the radius $r$ corresponding to the maximum of the potential shifts closer to the event horizon. This shift suggests that the presence of charge enhances the black hole's ability to trap perturbations near the horizon, leading to higher QNM frequencies. The charge $Q$ thus plays a dual role in both increasing the potential barrier and reducing the effective radius where the perturbation is concentrated, which may result in a more complex QNM spectrum.

We have also plotted similar curves for the electromagnetic potential in Figs. \ref{fig_Vem_01} and \ref{fig_Vem_02}, considering different values of the model parameters and multipole moments. The qualitative behavior of the electromagnetic potential curves is similar to that observed for the scalar potential. However, a notable difference is that the potential values for the scalar perturbation are generally higher than those for the electromagnetic perturbation. This difference suggests that the scalar field experiences a stronger interaction with the black hole spacetime, leading to higher potential barriers. Consequently, we expect the QNM spectrum for scalar perturbations to have higher frequencies and shorter lifetimes compared to electromagnetic perturbations. The lower potential values for the electromagnetic case imply that the QNMs associated with electromagnetic perturbations might exhibit lower frequencies and longer damping times, indicating a more gradual decay of these modes.

The analysis of the perturbation potential provides valuable insights into the nature of the QNMs for both scalar and electromagnetic perturbations. The variations in the potential with respect to parameters such as the multipole moment $l$, the model parameter $\alpha$, the coupling parameter $c_{\rm O}$, and the black hole charge $Q$ reveal how sensitive the QNMs are to these factors. This understanding is crucial for predicting the stability and dynamical response of black holes to perturbations, particularly in the context of different gravity models or in the presence of additional fields. One may note that with $Q=0$, one can easily obtain the corresponding potentials discussed in Ref. \cite{bh5}. However, we have noticed that the charge of the black hole i.e. $Q$ has a noticeable impact on the QNMs spectrum as predicted by the variation of the corresponding potential. In the presence of $Q$, the peak value of the potential increases and the variation is not similar to that of the Reissner-Nordström black hole. Hence, the QNMs spectrum seems to be different from that of a standard Reissner–Nordström black hole.

\subsection{Time-domain profile and quasinormal modes}

 To begin, let us first examine the time-domain profile of the scalar field before delving into the determination of QNMs. In the preceding section, we undertook extensive numerical computations to characterize the QNMs, thoroughly analyzing how these modes depend on the model parameters $Q$, $\alpha$, and $c_{\rm O}$. Our approach involved solving the perturbation equations in the frequency domain, which provided insights into the oscillatory behavior and damping times of the QNMs. However, frequency-domain analysis alone does not capture the full temporal evolution of the perturbations. Thus, in the subsequent section, our focus shifts towards understanding the time-domain behavior of both scalar and electromagnetic perturbations. This involves generating temporal profiles that illustrate how these fields evolve with time after an initial disturbance. To achieve this, we employ a time-domain integration framework based on the methodology proposed by Gundlach et al. \cite{gundlach}, which is well-suited for capturing the full dynamical evolution of perturbations. By applying this technique, we can observe the decay patterns and the late-time behavior of the perturbations, offering a complementary perspective to the frequency-domain analysis. This dual approach not only helps in verifying the consistency of our results but also provides a more comprehensive understanding of the dynamical properties of the perturbations in the given gravitational background. 

Now, to proceed with the time domain investigation, we at first define $\psi(r_*,
t) = \psi(i \Delta r_*, j \Delta t) = \psi_{i,j} $ and $V(r(r_*)) = V(r_*,t) =
V_{i,j}$. 
Using this expression, we write
\begin{equation}
\dfrac{\psi_{i+1,j} - 2\psi_{i,j} + \psi_{i-1,j}}{\Delta r_*^2} - \dfrac{%
\psi_{i,j+1} - 2\psi_{i,j} + \psi_{i,j-1}}{\Delta t^2} - V_i\psi_{i,j} = 0.
\end{equation}
In this scheme, we use the inittial conditions $\psi(r_*,t) = \exp \left[ -\dfrac{(r_*-k_1)^2%
}{2\sigma^2} \right]$ and $\psi(r_*,t)\vert_{t<0} = 0$ (note that $k_1$ and $%
\sigma$ are the median and width of the initial wave-packet in the analysis scheme). 
By utilising these expressions, one can obtain the time evolution of the scalar perturbation as shown below: 
\begin{equation}
\psi_{i,j+1} = -\,\psi_{i, j-1} + \left( \dfrac{\Delta t}{\Delta r_*}
\right)^2 (\psi_{i+1, j} + \psi_{i-1, j}) + \left( 2-2\left( \dfrac{\Delta t%
}{\Delta r_*} \right)^2 - V_i \Delta t^2 \right) \psi_{i,j}.
\end{equation}

One can obtain the required time domain profiles numerically by using this iteration scheme along with a suitable fixed value of the fraction 
$\frac{\Delta t}{\Delta r_*}$. Here a suitable value implies that the ratio $\frac{\Delta t}{\Delta r_*}$ must be less than $1$ in order to satisfy the Von Neumann stability condition.

\begin{figure}[htbp]
\centerline{
   \includegraphics[scale = 0.8]{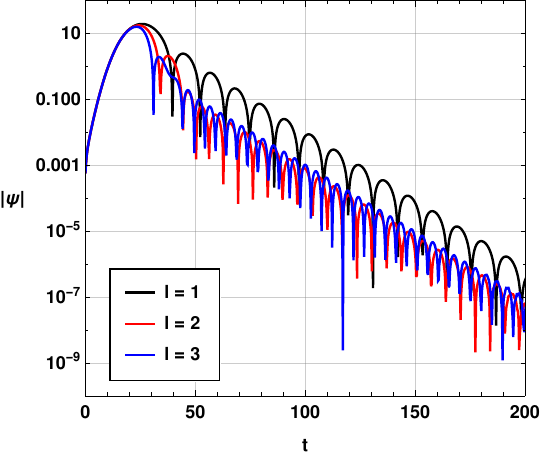}\hspace{0.5cm}
   \includegraphics[scale = 0.8]{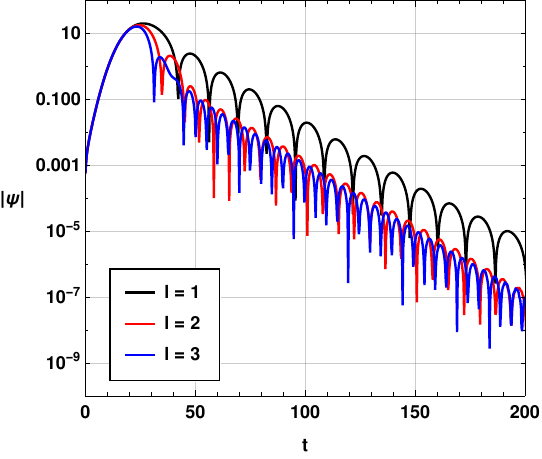}} \vspace{-0.2cm}
\caption{The time-domain profiles of the massless scalar
perturbations (first panel) and electromagnetic perturbations (right panel) for different multipole moments $l$ with the parameter values $M=1$, $G = 1$, $n= 0$, $\alpha =0.9$, $c_{\rm O}= 0.4$ and $Q = 0.3$. }
\label{time01}
\end{figure}

\begin{figure}[htbp]
\centerline{
   \includegraphics[scale = 0.8]{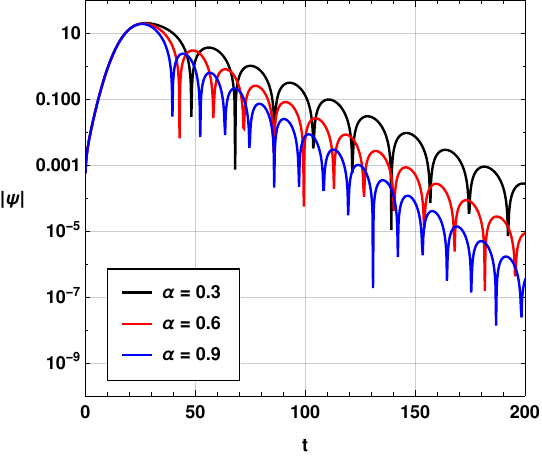}\hspace{0.5cm}
   \includegraphics[scale = 0.8]{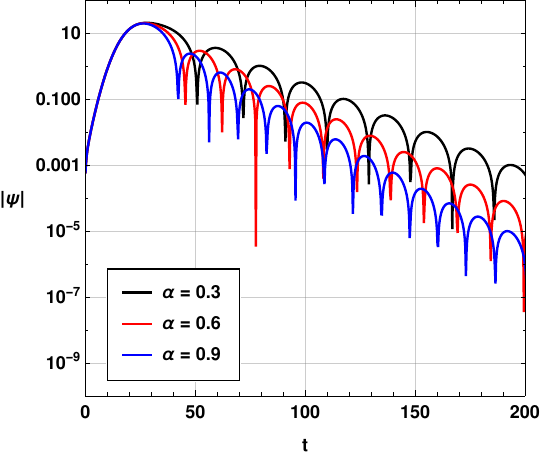}} \vspace{-0.2cm}
\caption{The time-domain profiles of the massless scalar
perturbations (first panel) and electromagnetic perturbations (right panel) for different values of the model parameter $\alpha$ with the parameter values $M=1$, $G = 1$, $l=1$, $n= 0,  c_{\rm O}= 0.4$ and $Q = 0.3$.}
\label{time02}
\end{figure}

\begin{figure}[htbp]
\centerline{
   \includegraphics[scale = 0.8]{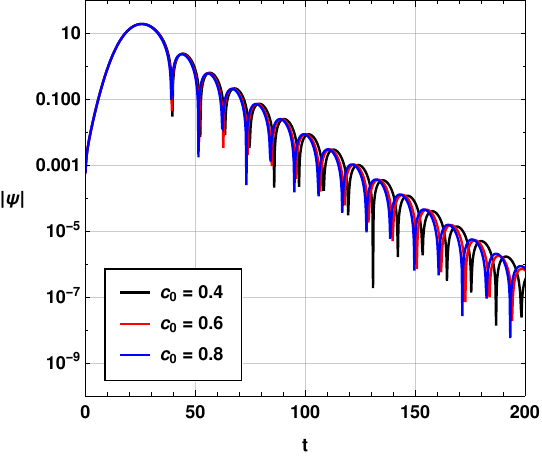}\hspace{0.5cm}
   \includegraphics[scale = 0.8]{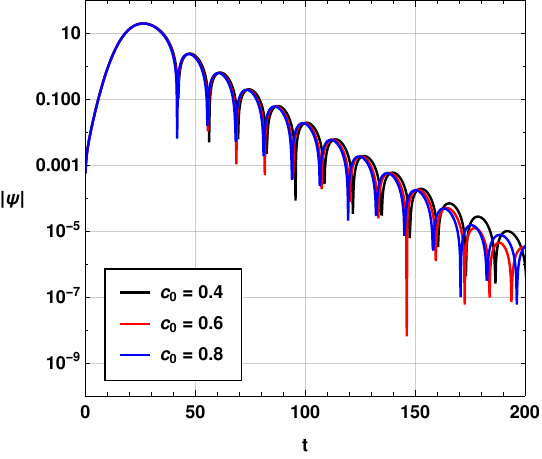}} \vspace{-0.2cm}
\caption{The time-domain profiles of the massless scalar
perturbations (first panel) and electromagnetic perturbations (right panel) for different values of $c_{\rm O}$ with the parameter values $M=1$, $G = 1$, $l=1$, $n= 0, \alpha = 0.9$ and $Q = 0.3$.}
\label{time03}
\end{figure}

\begin{figure}[htbp]
\centerline{
   \includegraphics[scale = 0.8]{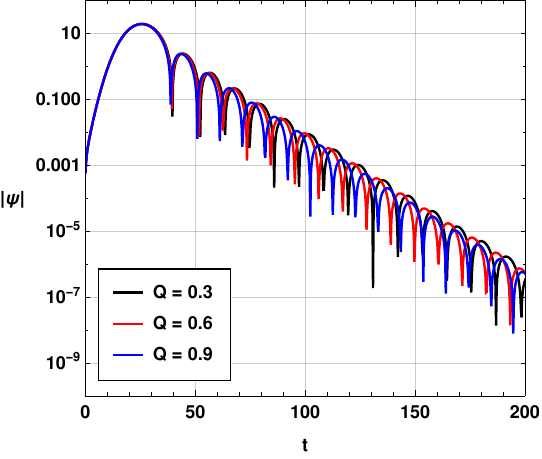}\hspace{0.5cm}
   \includegraphics[scale = 0.8]{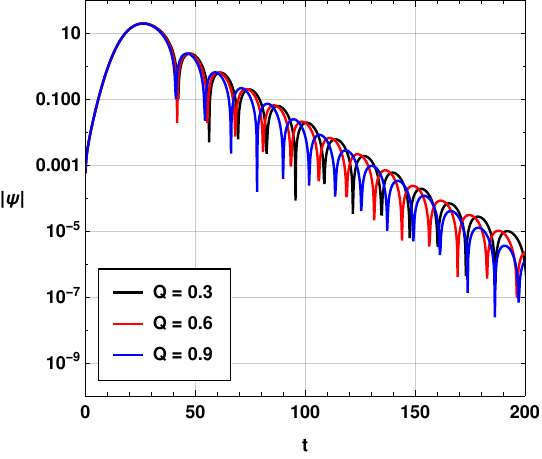}} \vspace{-0.2cm}
\caption{The time-domain profiles of the massless scalar
perturbations (first panel) and electromagnetic perturbations (right panel) for different values of $Q$ with the parameter values $M=1$, $G = 1$, $l=1$, $n= 0, \alpha = 0.9$ and $c_{\rm O}= 0.4$.}
\label{time04}
\end{figure}

The time-domain profiles of scalar and electromagnetic perturbations provide important insights into the dynamic behavior of these fields in black hole spacetimes. The left panel of Fig. \ref{time01} presents the temporal profile of scalar perturbations, while the right panel shows that of electromagnetic perturbations. The chosen parameters include an overtone number of $n = 0$, with $\alpha = 0.9$, $Q = 0.3$, and $c_{\rm O} = 0.4$, while the multipole moment $l$ is varied. Both profiles reveal that an increase in $l$ leads to higher frequencies of oscillations. However, the decay rates behave differently for scalar and electromagnetic perturbations. In the case of scalar perturbations, the decay rate increases significantly as $l$ grows, meaning that higher multipole moments correspond to faster damping. In contrast, the decay rate for electromagnetic perturbations shows only a slight variation with increasing $l$. Additionally, scalar perturbations tend to damp out faster than electromagnetic ones, implying that scalar fields are more sensitive to perturbations and exhibit quicker stabilization in the black hole environment.

Fig. \ref{time02} further explores the time-domain profiles by varying the model parameter $\alpha$, keeping the overtone number $n=0$, multipole moment $l=1$, $Q=0.3$, and $c_{\rm O} = 0.4$ fixed. Similar to the previous case, the decay rate for scalar perturbations is consistently greater than that for electromagnetic perturbations. This suggests that $\alpha$, which is likely a characteristic of the black hole or the surrounding spacetime, exerts a more pronounced influence on the scalar field. As the value of $\alpha$ increases, both the oscillation frequency and decay rate are affected, but the damping of scalar perturbations remains more rapid than that of electromagnetic perturbations. The scalar field tends to dissipate faster, reflecting its stronger interaction with the black hole geometry compared to electromagnetic fields.

Fig. \ref{time03} examines the effects of the parameter $c_{\rm O}$ on the time-domain profiles. Here, for small values of time $t$, both scalar and electromagnetic perturbations exhibit very similar profiles, with minimal differences in their oscillation patterns. However, as time progresses, variations in oscillation frequencies become more noticeable, particularly for larger values of $t$. Despite these differences, the damping rates for both types of perturbations appear relatively unaffected by changes in $c_{\rm O}$, suggesting that this parameter influences the frequency of oscillations more than the rate at which the perturbations decay. For the range of $c_{\rm O}$ values considered, the impact on damping seems marginal, indicating that $c_{\rm O}$ primarily affects the oscillatory dynamics rather than the overall stability.

Finally, in Fig. \ref{time04}, the time-domain profiles are plotted with varying values of the charge parameter $Q$. The results indicate that the black hole charge has a noticeable effect on the oscillation frequency for both scalar and electromagnetic perturbations. As $Q$ increases, the oscillation frequencies increase, suggesting that a more highly charged black hole causes perturbations to oscillate more rapidly. The damping rate, however, increases only slightly for both types of perturbations, implying that while the charge of the black hole influences the oscillatory behavior, it has a relatively minor impact on the overall rate at which the perturbations decay. This finding highlights that the black hole charge primarily affects the dynamic response of the system without drastically altering its stability over time.

The time-domain profiles of scalar and electromagnetic perturbations reveal key differences in how these fields evolve over time in the presence of a black hole. Scalar perturbations generally exhibit higher decay rates and faster damping than electromagnetic perturbations, making them more responsive to changes in the black hole's parameters, such as the multipole moment $l$, the model parameter $\alpha$, and the charge $Q$. These results provide a deeper understanding of the dynamics of black hole perturbations and the influence of various physical parameters on the behavior of QNMs.

\subsection{WKB method with Pad\'e Approximation for Quasinormal modes}

In this study, we aim to estimate the QNMs of the black holes by employing the Wentzel-Kramers-Brillouin (WKB) method, a widely recognized approximation technique in the field of black hole perturbation theory and investigate the validity of Hod's conjecture. Initially introduced by Schutz and Will \cite{Schutz}, the WKB method provides a first-order approximation for calculating QNMs. However, despite its utility, the method is known to exhibit certain limitations, including a higher degree of error in some cases. To address these limitations, researchers have developed higher-order WKB approximations, significantly improving the accuracy of QNM calculations \cite{Will_wkb, Konoplya_wkb, Maty_wkb}. Our work utilizes the advancements suggested by Ref. \cite{Maty_wkb}, where Pad\'e approximations are incorporated into the WKB method, yielding more precise results.

Building on these improvements, we apply the Pad\'e-averaged 6th-order WKB approximation technique to estimate the QNMs of black holes. The incorporation of Pad\'e averaging has been shown to enhance the precision of QNM estimates, as demonstrated in prior studies such as Ref. \cite{Konoplya_wkb}. By using this refined approach, we seek to achieve more accurate and reliable estimates of QNMs, which will be compared with earlier results to validate our findings. This comparative analysis is critical in evaluating the robustness of the WKB method and advancing our understanding of black hole dynamics.

In this subsection, we present QNM values computed using the Pad\'e-averaged 6th-order WKB approximation, along with error estimates in Tables \ref{tab01} and \ref{tab02}. The 3rd and 4th columns of these tables show the errors associated with the WKB approximation, specifically the root mean square (rms) error, denoted as $\vartriangle_{rms}$, and the error term $\Delta_6$, defined as \cite{Konoplya_wkb}
\begin{equation}
\Delta_6 = \dfrac{\vline \; \omega_7 - \omega_5 \; \vline}{2},
\end{equation}
where $\omega_7$ and $\omega_5$ represent the QNMs computed using the 7th and 5th-order Padé-averaged WKB methods, respectively. By analyzing these errors, we provide a detailed assessment of the accuracy of our QNM estimates, further reinforcing the reliability of the WKB method in black hole perturbation analysis.

\begin{table}[ht]
\caption{The scalar QNMs of the CSBH for $n= 0$, $M=1$, $G = 1$, $\alpha = 0.9$, $Q=0.5$ and $c_{\rm O}=0.3$ using the 6th order WKB approximation method averaged with  Pad\'e approximants.}
\label{tab01}
\begin{center}
{\small 
\begin{tabular}{|cccc|}
\hline
\;\;$l$ &  \;\; Pad\'e averaged WKB\;\;
& $\vartriangle_{rms}$ & $\Delta_6$ \\ \hline
$l=1$ & $0.27937 - 0.094328i$ & $4.74121\times 10^{-6}$
& $0.0000176752$  \\ 
$l=2$ & $0.464267 - 0.0924163i$  & $%
6.44457\times10^{-7}$ & $2.44832\times10^{-6}$ \\ 
$l=3$ & $0.649716 - 0.0918554i$  & $%
5.0719\times10^{-8}$ & $4.46484\times10^{-7}$ \\ 
$l=4$ & $0.835258 - 0.0916204i$ & $%
8.56836\times10^{-9}$ & $1.14886\times10^{-7}$  \\ 
$l=5$ & $1.02083 - 0.0915005i$  & $%
2.25924\times10^{-9}$ & $4.54218\times10^{-8}$\\ \hline
\end{tabular}
}
\end{center}
\end{table}

\begin{table}[ht]
\caption{The electromagnetic QNMs of the CSBH  for $n= 0$, $M=1$, $G = 1$, $\alpha = 0.9$, $Q=0.5$ and $c_{\rm O}=0.3$ using the 6th order WKB approximation method averaged with  Pad\'e approximants.}
\label{tab02}
\begin{center}
{\small 
\begin{tabular}{|cccc|}
\hline
\;\;$l$ &  \;\; Pad\'e averaged WKB\;\;
& $\vartriangle_{rms}$ & $\Delta_6$ \\ \hline
$l=1$ & $0.24098 - 0.0880597i$ & $3.87696\times 10^{-6}$
& $0.0000234139$  \\ 
$l=2$ & $0.442189 - 0.090206i$  & $%
1.83268\times10^{-7}$ & $2.32131\times10^{-6}$ \\ 
$l=3$ & $0.634127 - 0.0907312i$  & $%
2.44681\times10^{-8}$ & $4.26197\times10^{-7}$ \\ 
$l=4$ & $0.82319 - 0.090941i$ & $%
5.37897\times10^{-9}$ & $1.09616\times10^{-7}$  \\ 
$l=5$ & $1.01098 - 0.0910459i$  & $%
1.60587\times10^{-9}$ & $4.17134\times10^{-8}$\\
\hline
\end{tabular}
}
\end{center}
\end{table}

In Table \ref{tab01}, the QNMs for massless scalar perturbations of a CSBH are presented, with the following model parameters: $ c_{\rm O} = 0.3 $, $ Q = 0.5 $, $ \alpha = 0.9 $, $ G = M = 1 $, and overtone number $ n = 0 $. The results are obtained using the 6th-order WKB approximation method, averaged with Pad\'e approximants. The table also shows two types of errors: the root mean square (rms) error ($ \vartriangle_{rms} $) and the relative error term $ \Delta_6 $. It is observed that while the Pad\'e-averaged WKB method tends to have larger relative errors ($ \Delta_6 $) for lower values of the multipole moment $ l $, the rms error ($ \vartriangle_{rms} $) is relatively smaller. As the multipole moment $ l $ increases, both error terms decrease significantly. This behavior is typical of the WKB method, which becomes less accurate when the difference between the multipole moment $ l $ and the overtone number $ n $ is small. Additionally, the quasinormal frequencies and damping rates increase as the value of $ l $ increases.

Table \ref{tab02} presents the QNMs for electromagnetic perturbations under the same model parameters and method used in Table \ref{tab01}. Similar to the case of scalar perturbations, the errors are larger for lower multipole moments $ l $, but as $ l $ increases, both the rms and relative errors decrease. This indicates that the accuracy of the WKB approximation improves with larger multipole moments, consistent with the results observed in scalar perturbations.

When comparing the results from both tables, it is clear that the quasinormal frequencies and damping rates for electromagnetic perturbations are lower than those for scalar perturbations. This difference highlights the impact of the type of perturbation on the black hole's quasinormal modes, where electromagnetic perturbations generally produce lower frequency oscillations and slower decay rates compared to massless scalar perturbations.

\begin{figure}[htbp]
\centerline{
   \includegraphics[scale = 0.5]{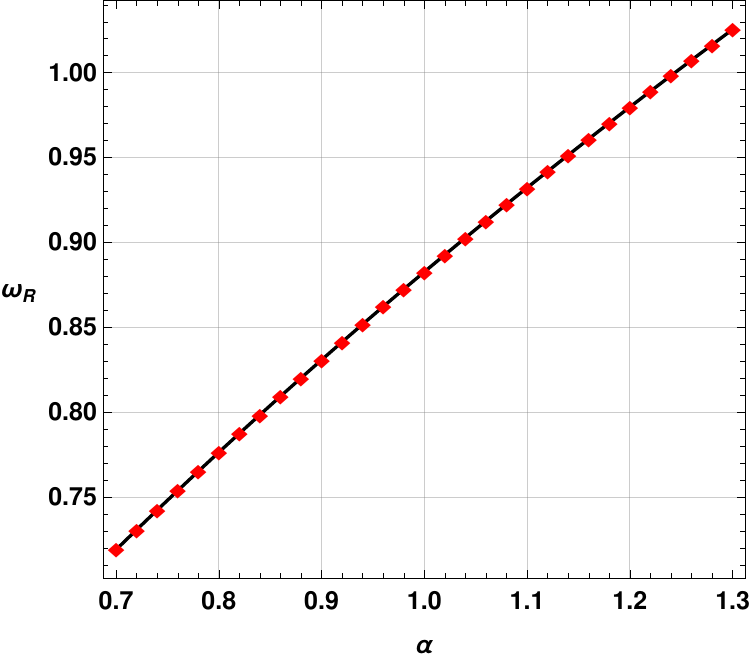}\hspace{0.5cm}
   \includegraphics[scale = 0.5]{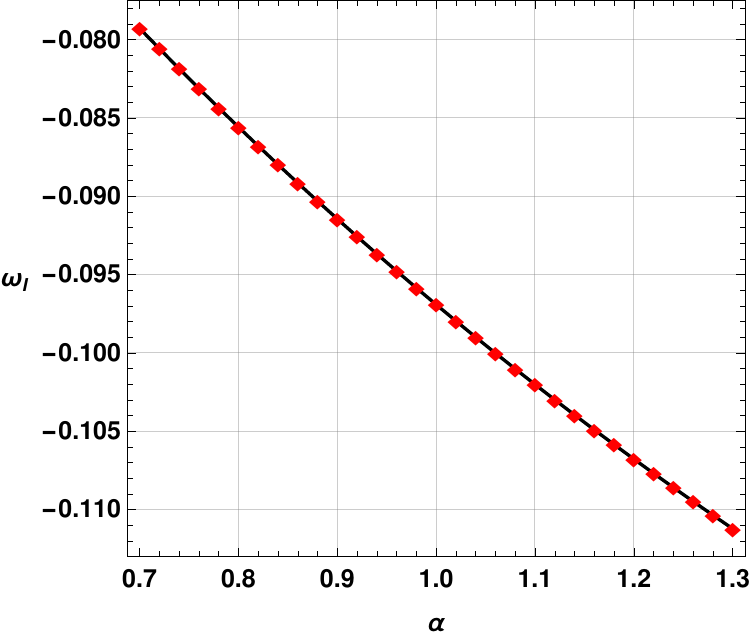}} \vspace{-0.2cm}
\caption{The real (left panel) and imaginary (right panel) parts of the QNMs of the CSBH for massless scalar perturbations as a function of the vacuum energy parameter $\alpha$ with $M=1$, $G = 1$, $n= 0$, $l=4$ and $c_{\rm O} = 0.4$.}
\label{QNMs01}
\end{figure}

\begin{figure}[htbp]
\centerline{
   \includegraphics[scale = 0.5]{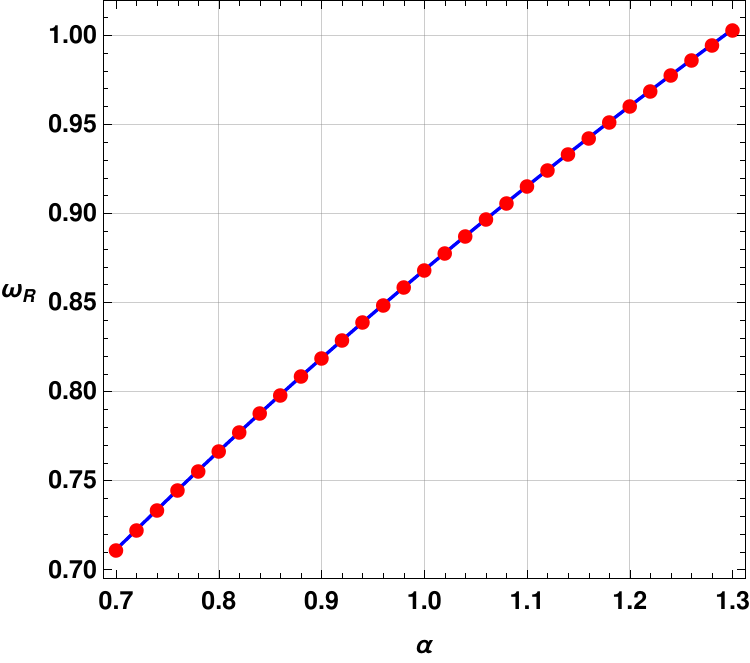}\hspace{0.5cm}
   \includegraphics[scale = 0.5]{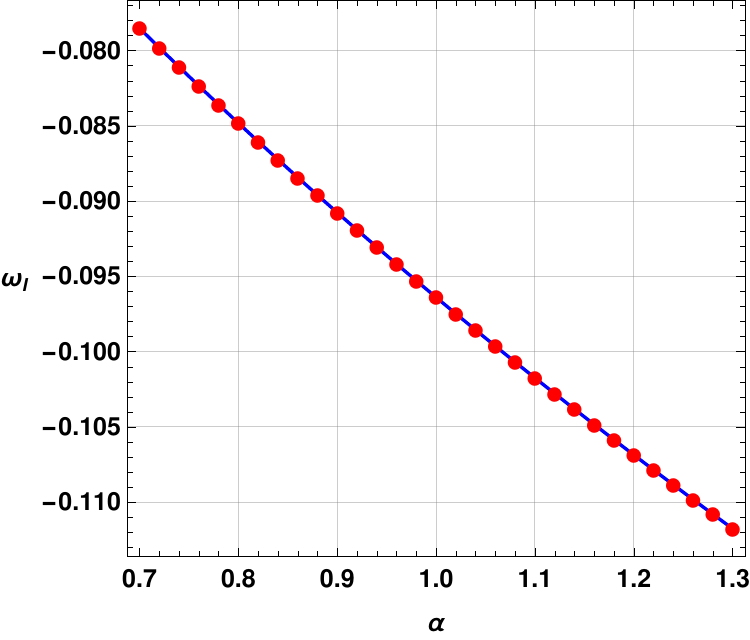}} \vspace{-0.2cm}
\caption{The real (left panel) and imaginary (right panel) parts of the QNMs of the CSBH for electromagnetic perturbations as a function of the vacuum energy parameter $\alpha$ with $M=1$, $G = 1$, $n= 0$, $l=4$ and $c_{\rm O} = 0.4$.}
\label{QNMs02}
\end{figure}

\begin{figure}[htbp]
\centerline{
   \includegraphics[scale = 0.5]{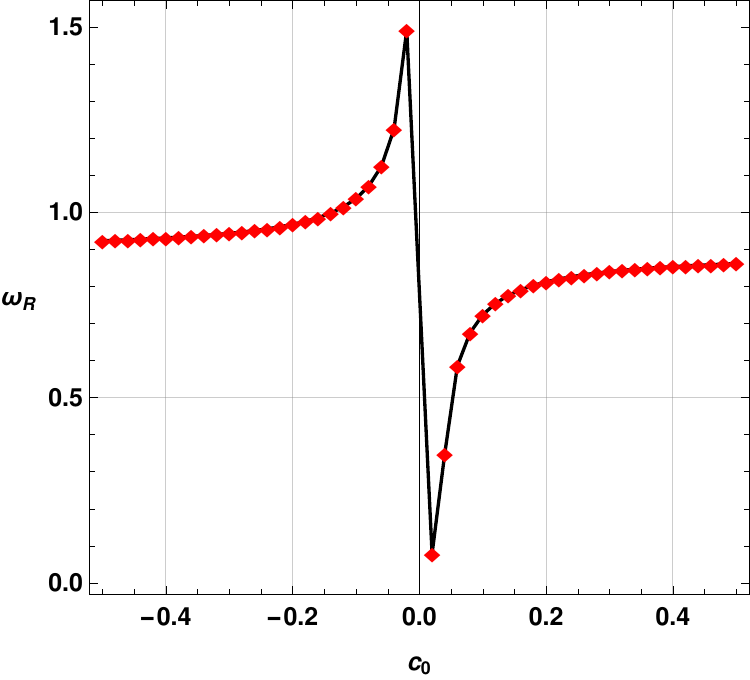}\hspace{0.5cm}
   \includegraphics[scale = 0.5]{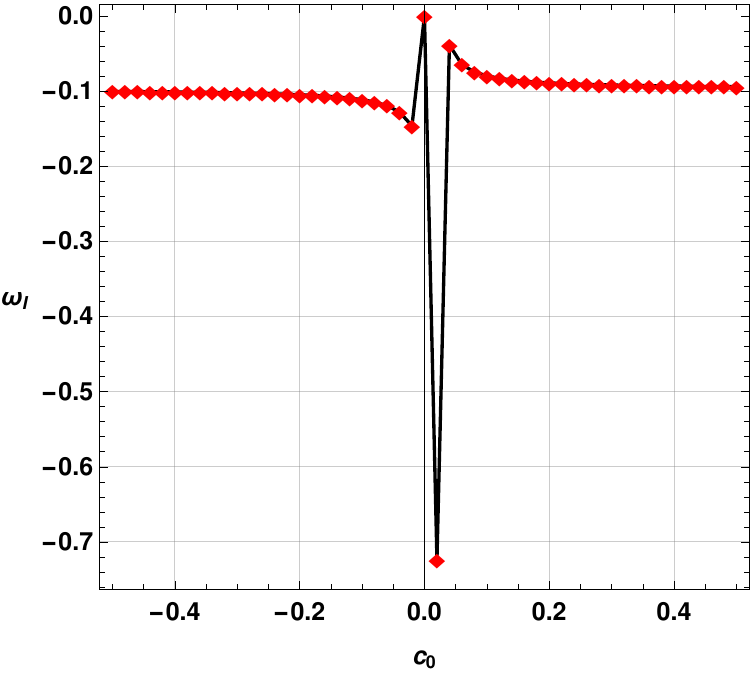}} \vspace{-0.2cm}
\caption{The real (left panel) and imaginary (right panel) parts of the QNMs of the CSBH for massless scalar 
perturbations as a function of the symmergent parameter $c_{\rm O}$ with $M=1$, $G = 1$, $n= 0$, $l=4$, $Q=0.55$ and $\alpha = 0.9$.}
\label{QNMs03}
\end{figure}

\begin{figure}[htbp]
\centerline{
   \includegraphics[scale = 0.5]{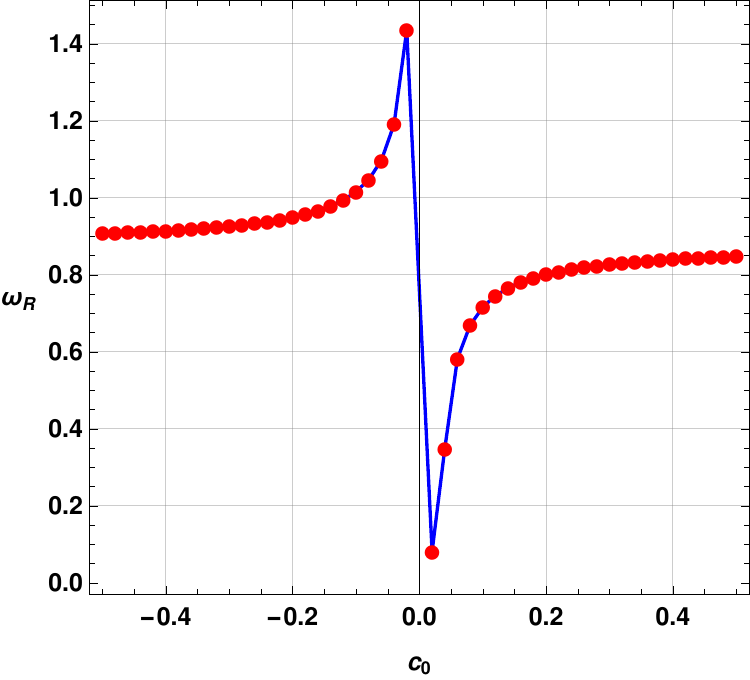}\hspace{0.5cm}
   \includegraphics[scale = 0.5]{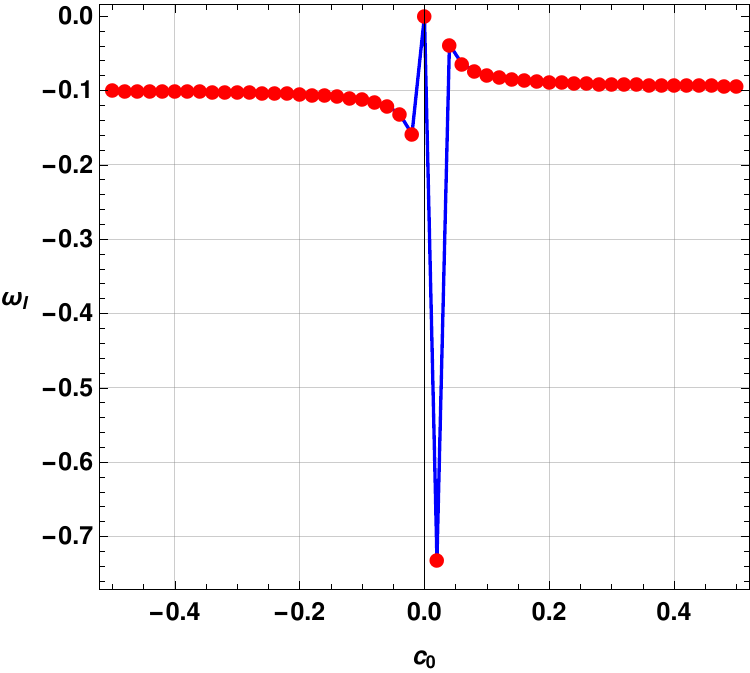}} \vspace{-0.2cm}
\caption{The real (left panel) and imaginary (right panel) parts of the QNMs of the CSBH for electromagnetic perturbations as a function of the symmergent parameter $c_{\rm O}$ with $M=1$, $G = 1$, $n= 0$, $l=4$ and $\alpha = 0.9$.}
\label{QNMs04}
\end{figure}

\begin{figure}[htbp]
\centerline{
   \includegraphics[scale = 0.5]{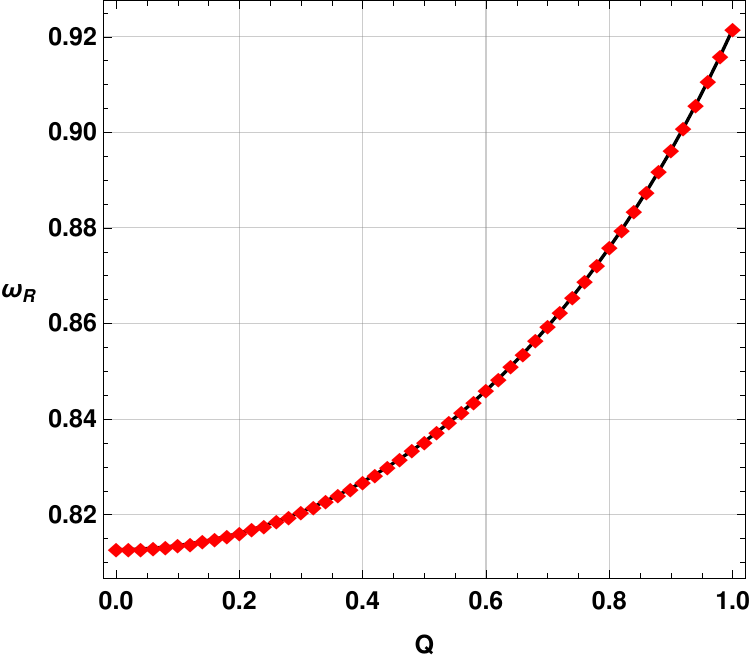}\hspace{0.5cm}
   \includegraphics[scale = 0.5]{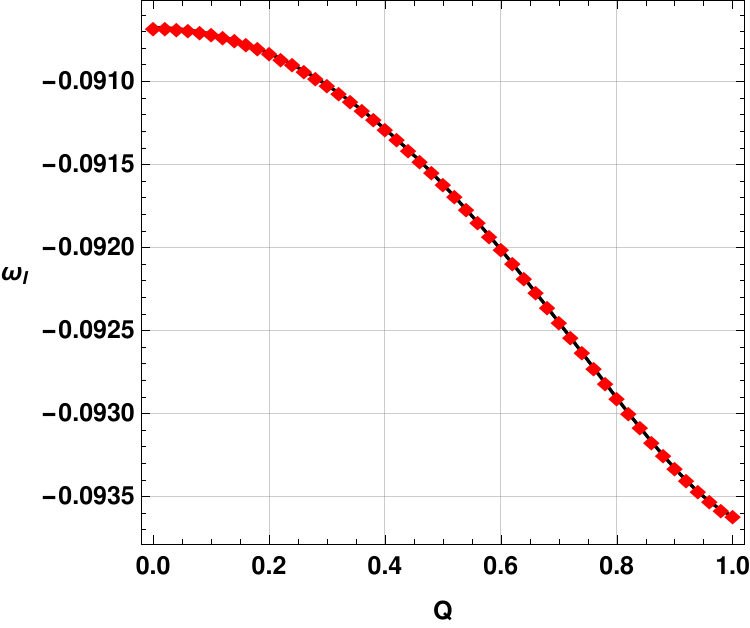}} \vspace{-0.2cm}
\caption{The real (left panel) and imaginary (right panel) parts of the QNMs of the CSBH for massless scalar perturbations as a function of the charge parameter $Q$ with $M=1$, $G = 1$, $n= 0$, $l=4$, $\alpha=0.9$ and $c_{\rm O}=0.3$. }
\label{QNMs05}
\end{figure}

\begin{figure}[htbp]
\centerline{
   \includegraphics[scale = 0.5]{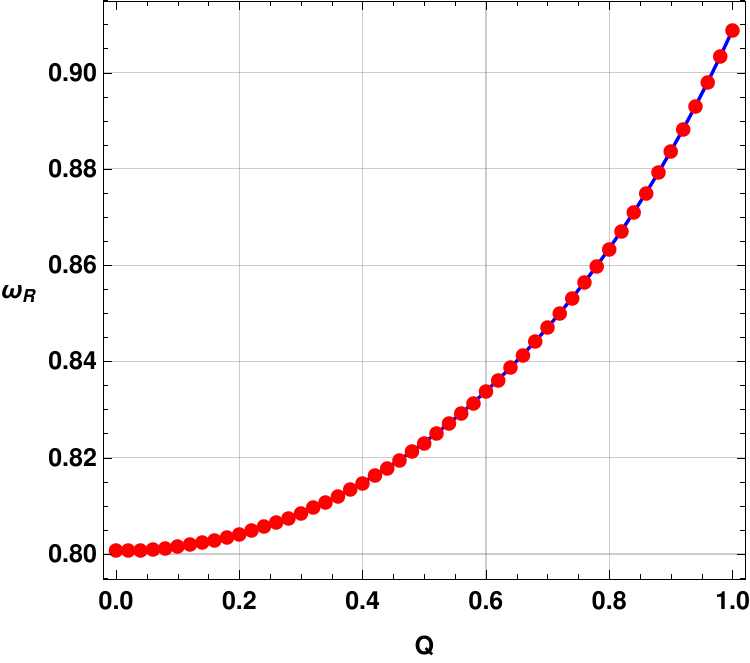}\hspace{0.5cm}
   \includegraphics[scale = 0.5]{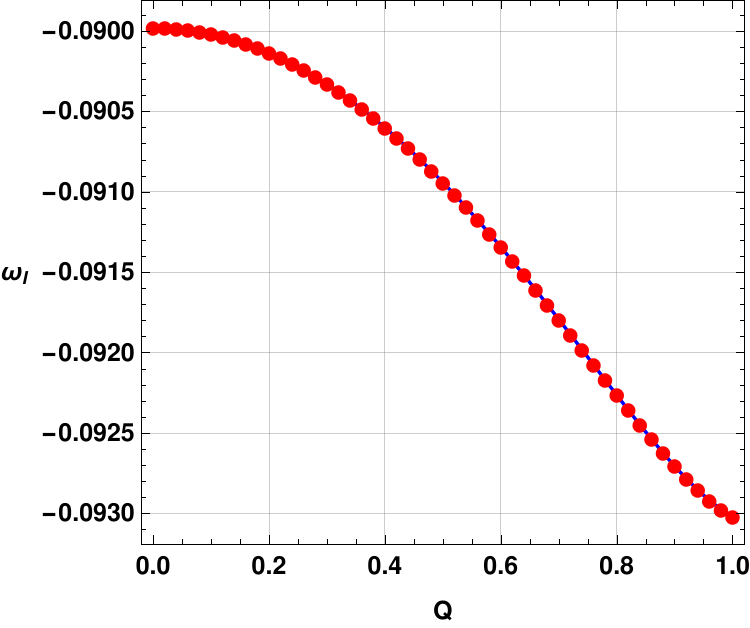}} \vspace{-0.2cm}
\caption{The real (left panel) and imaginary (right panel) parts of the QNMs of the CSBH for electromagnetic perturbations as a function of the charge parameter $Q$ with $M=1$, $G = 1$, $n= 0$, $l=4$, $\alpha=0.9$ and $c_{\rm O}=0.3$.}
\label{QNMs06}
\end{figure}

\begin{figure}[htbp]
\centering
   \includegraphics[scale=0.35]{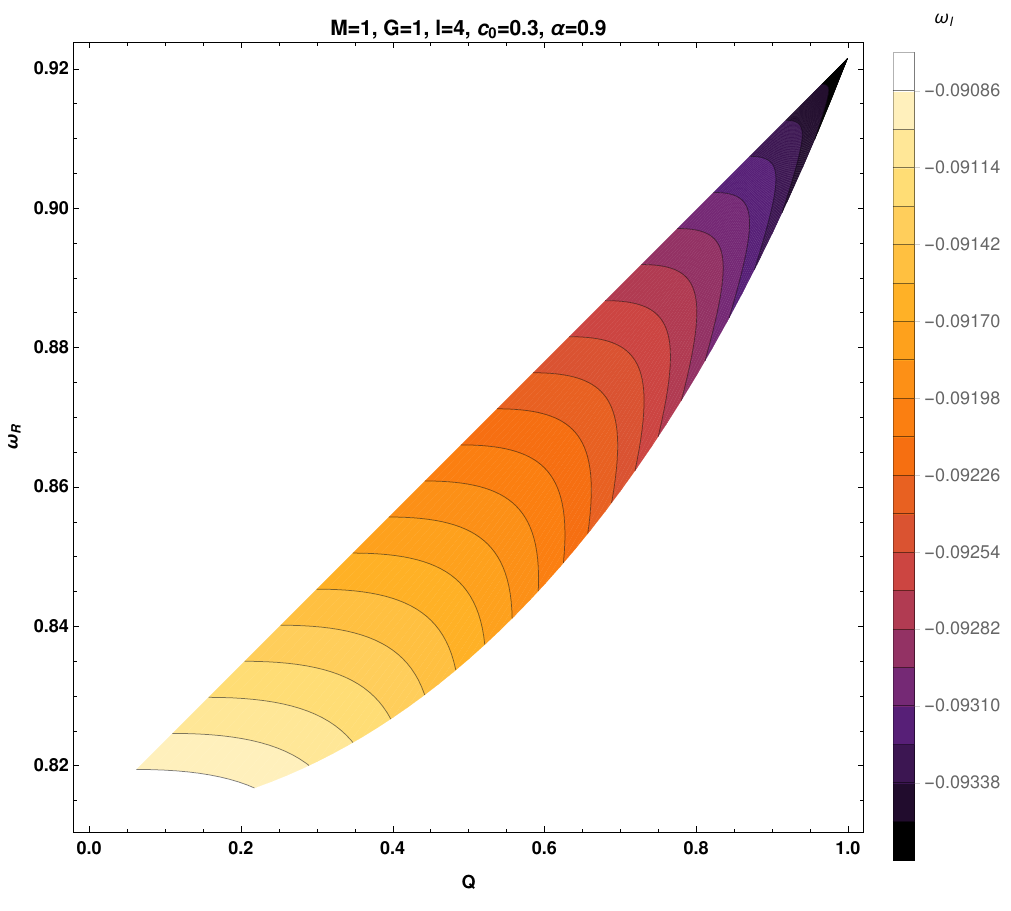} \hspace{0.5cm}
   \includegraphics[scale=0.35]{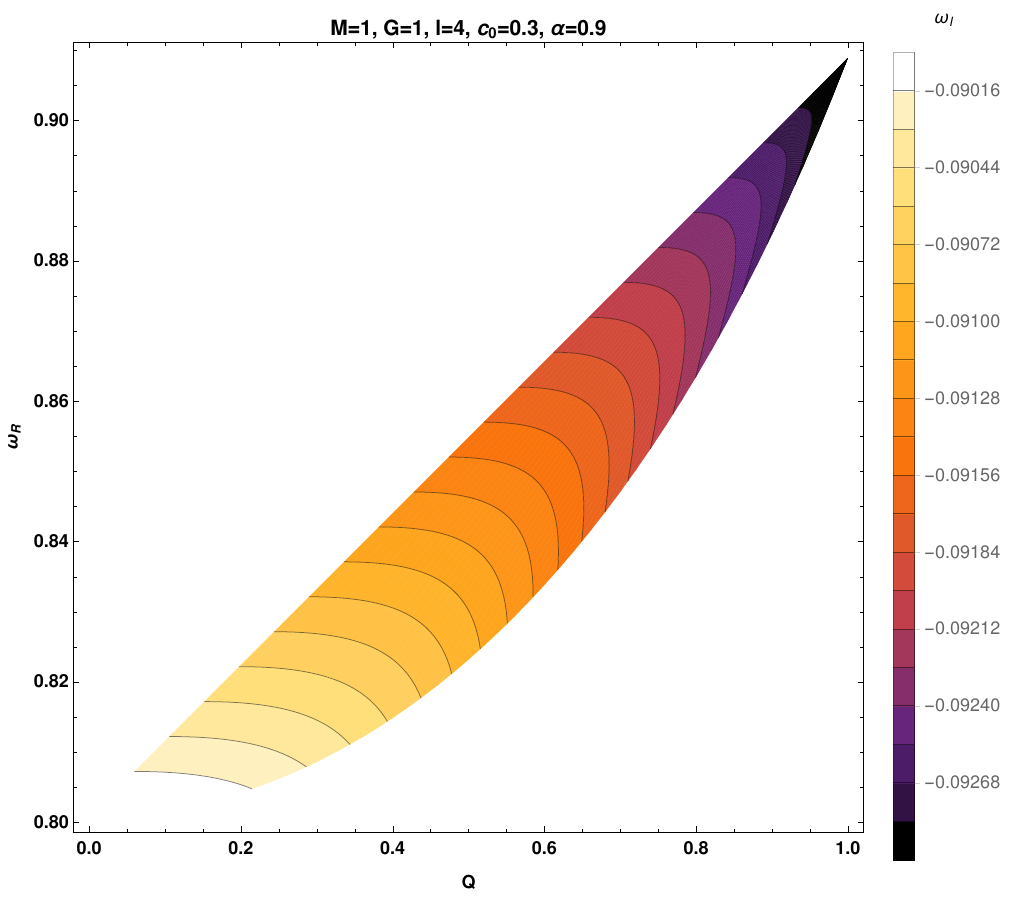} \hspace{0.5cm}
   \includegraphics[scale=0.35]{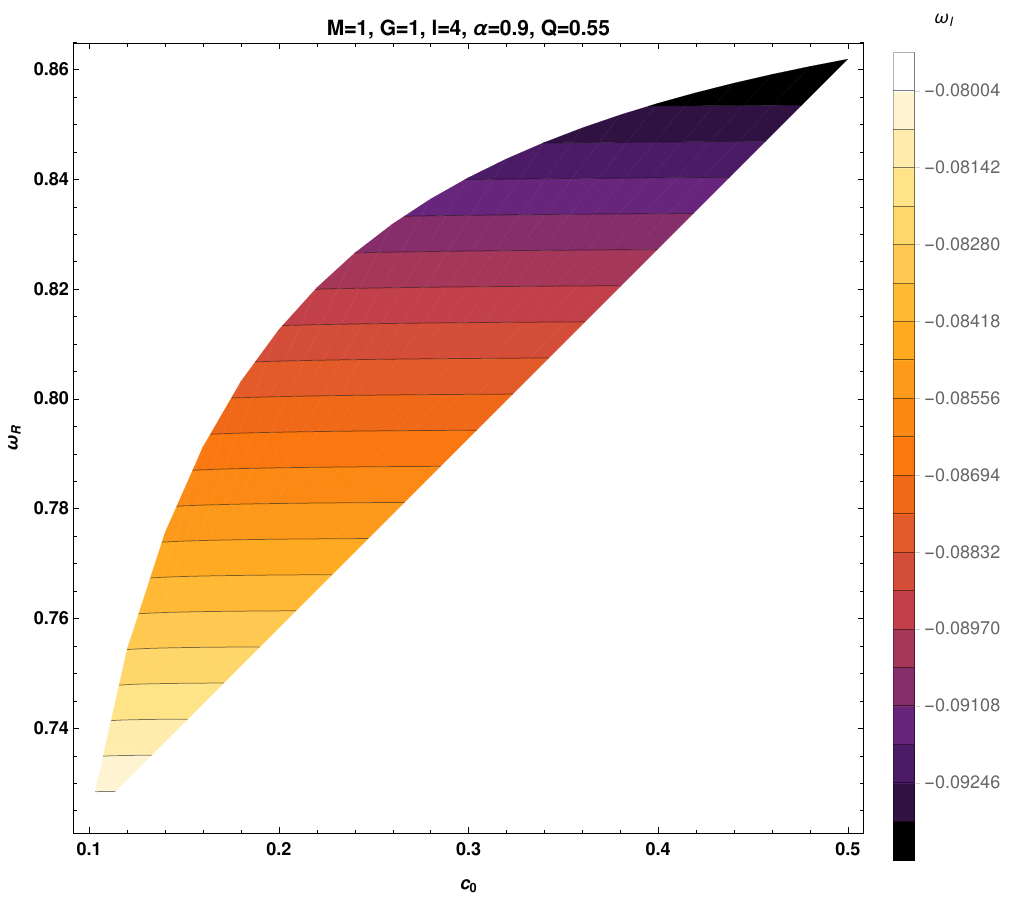} \hspace{0.5cm}
   \includegraphics[scale=0.35]{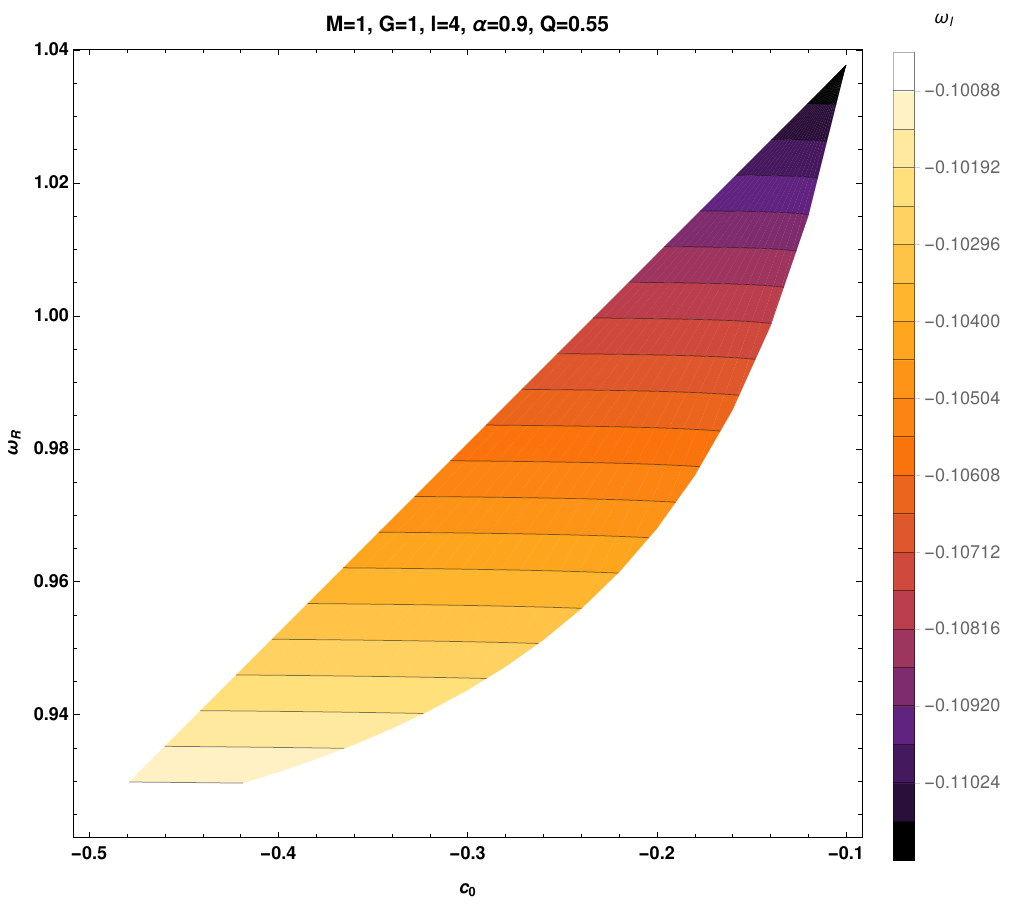} \hspace{0.5cm}
   \includegraphics[scale=0.35]{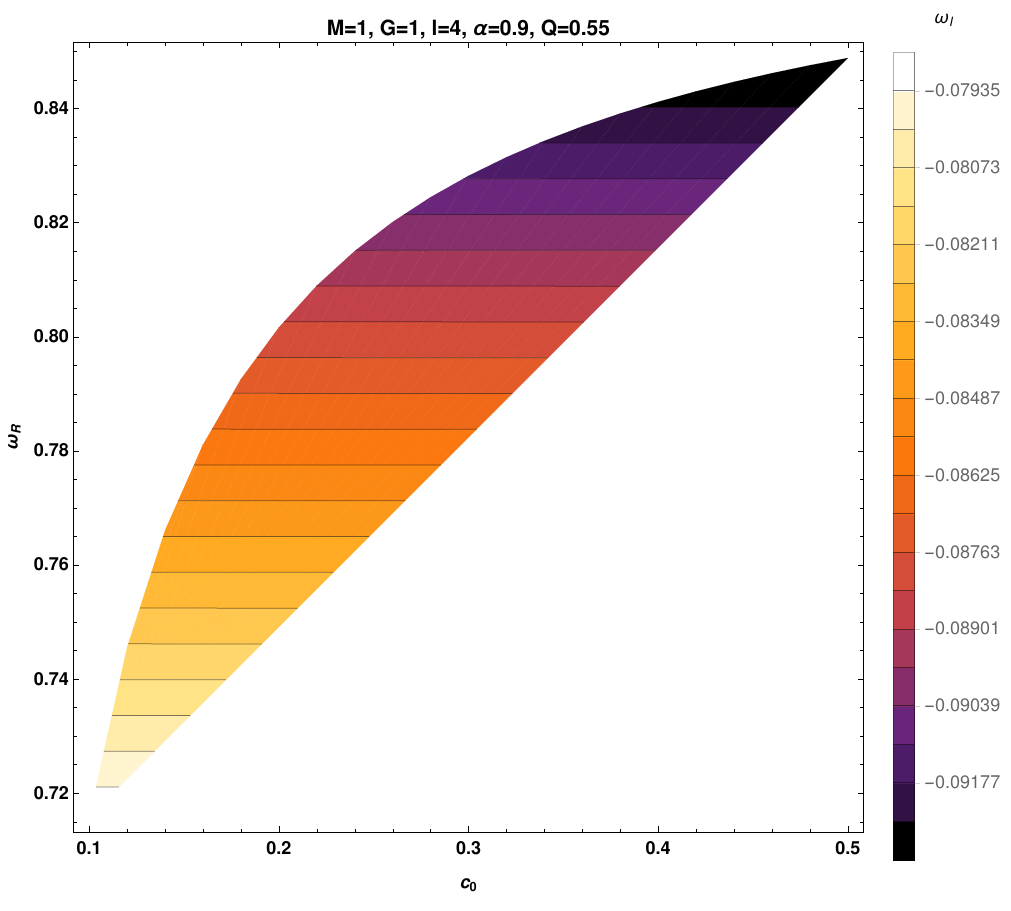} \hspace{0.5cm}
   \includegraphics[scale=0.35]{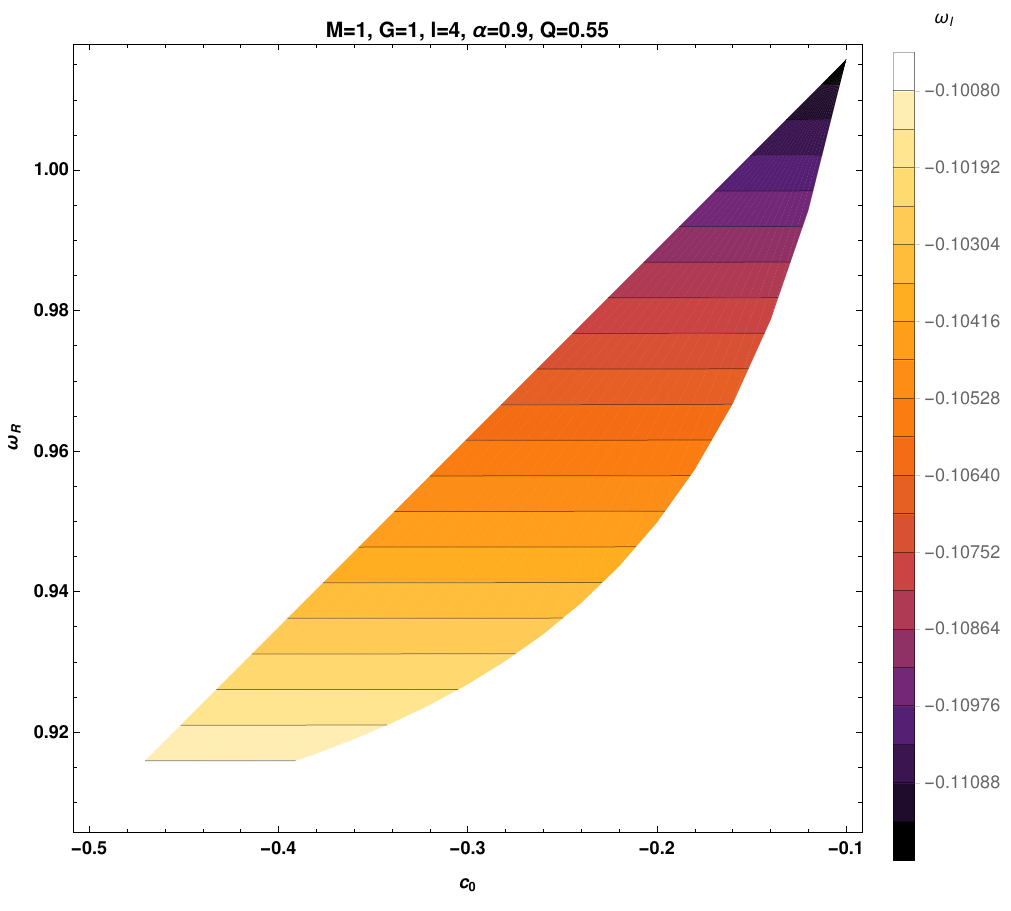} \\
   \includegraphics[scale=0.35]{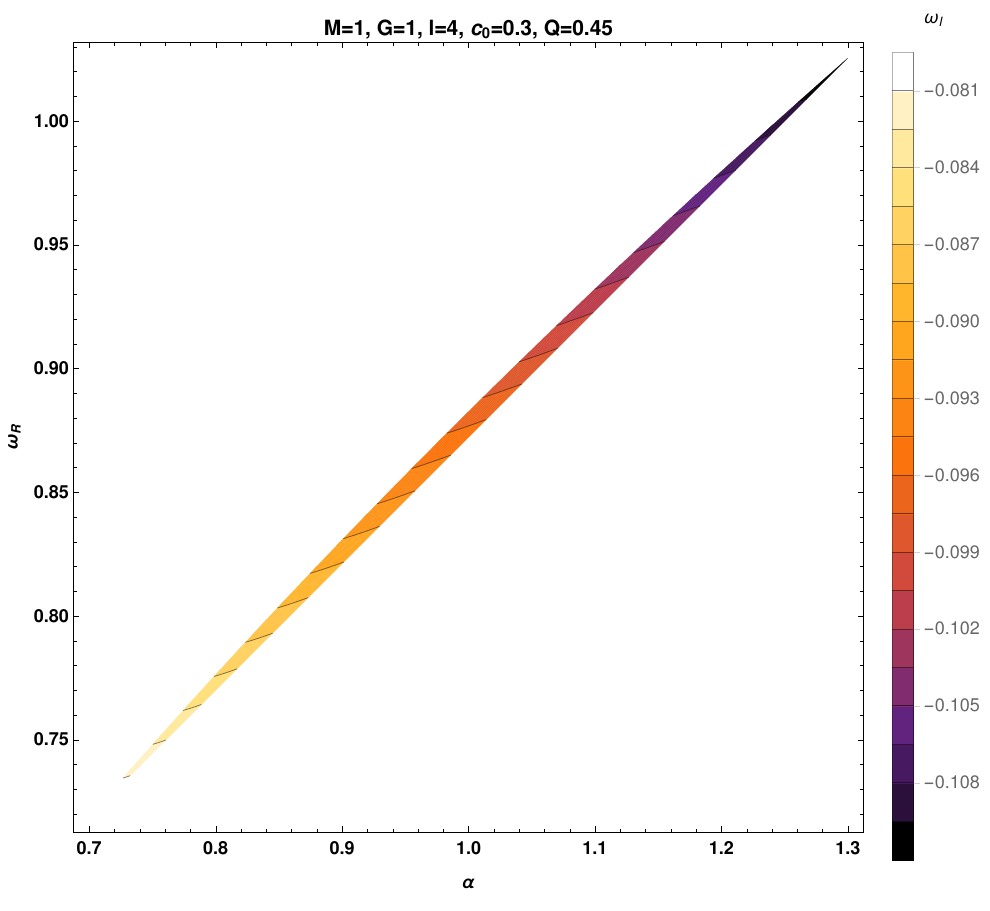} \hspace{0.5cm}
   \includegraphics[scale=0.35]{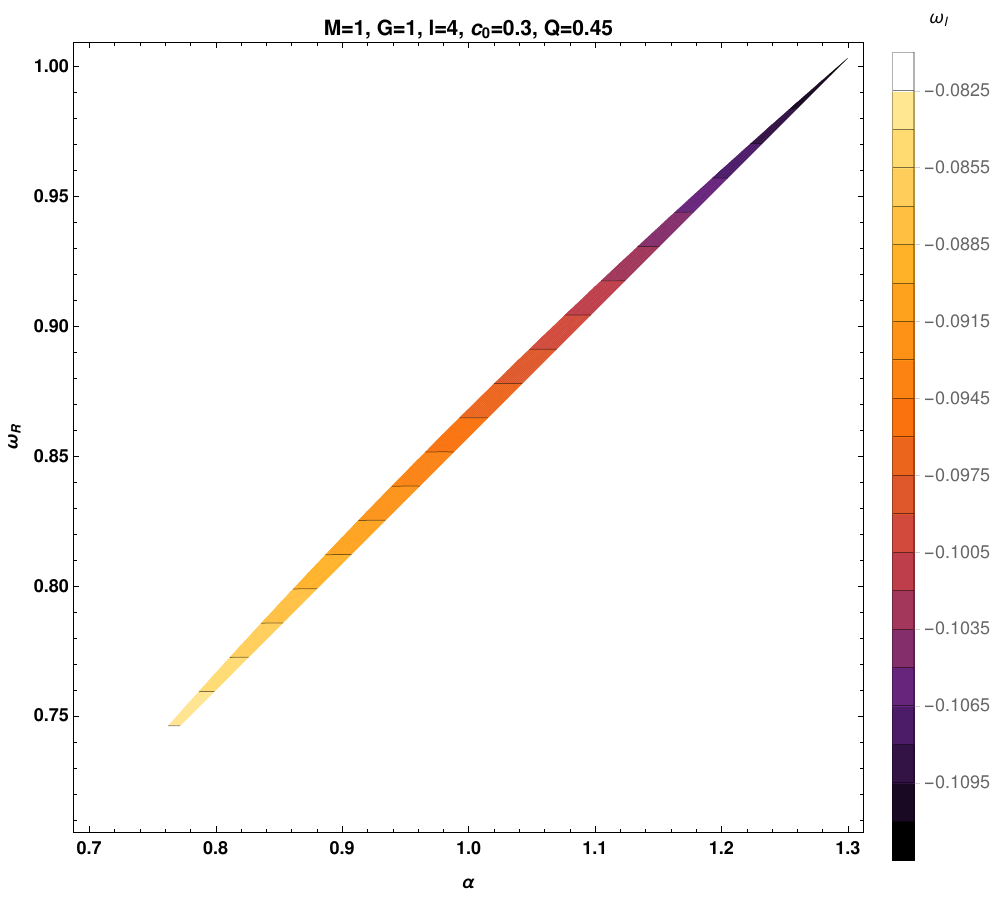}
\vspace{-0.2cm}
\caption{Contours showing the variation of QNMs with model parameters. First column shows the scalar QNMs and second column shows the electromagnetic QNMs.}
\label{QNMs07}
\end{figure}

To explore the effects of the model parameters on the QNM spectrum, we have explicitly plotted the real and imaginary components of the QNMs with respect to these parameters. For this purpose, we employed the Padé-averaged 6th-order WKB approximation method, choosing a higher value of the multipole moment $ l $. The rationale for selecting a larger value of $ l $ is that the WKB method's associated error diminishes significantly at higher values of $ l $. In Fig. \ref{QNMs01}, the variation of real (left) and imaginary (right) QNMs with respect to the model parameter $ \alpha $ is shown for scalar perturbations. As $ \alpha $ increases, the real QNMs, which correspond to the oscillation frequencies, exhibit a significant increase, following an almost linear pattern. It's important to note that $ \alpha < 1 $ implies a de Sitter (dS) spacetime, meaning that de Sitter black holes have lower oscillation frequencies compared to anti-de Sitter (AdS) or asymptotically flat black holes. The damping rate or decay rate of GWs also increases linearly with $ \alpha $. The same behavior is observed in Fig. \ref{QNMs02} for electromagnetic perturbations.

In Fig. \ref{QNMs03}, the variation of the ringdown GW frequency and damping rate with respect to the parameter $ c_{\rm O} $ is depicted for scalar perturbations. For positive values of $ c_{\rm O} $, both real quasinormal frequencies and damping rates increase non-linearly, with the variation being more pronounced for smaller values of $ c_{\rm O} $. As $ c_{\rm O} $ approaches 0.4, the variation in both real and imaginary QNMs becomes negligible. In contrast, for negative values of $ c_{\rm O} $, both the oscillation frequency and damping rate increase non-linearly, reaching a maximum near $ c_{\rm O} = 0 $. Notably, a discontinuity in the QNM spectrum is observed at $ c_{\rm O} = 0 $. Fig. \ref{QNMs04} presents similar results for electromagnetic perturbations, showing comparable behavior across the parameter space.

The charge parameter $ Q $ has a significant and nonlinear impact on the QNM spectrum of black holes. As shown in Figs. \ref{QNMs05} and \ref{QNMs06}, both the oscillation frequency (real part of QNMs) and the damping rate (imaginary part of QNMs) increase as $ Q $ increases. This behavior holds true for both scalar and electromagnetic perturbations, although the precise values of the frequencies and decay rates differ depending on the type of perturbation. In the case of electromagnetic perturbations, the oscillation frequencies and damping rates are generally lower than those observed for scalar perturbations. The nonlinear nature of the relationship between $ Q $ and the QNMs suggests that even small changes in the charge parameter can have a pronounced effect on the behavior of the black hole.

Physically, the increase in oscillation frequency with higher $ Q $ can be understood in the context of the black hole's enhanced electric field. As the black hole’s charge increases, the strength of its electromagnetic field grows, which leads to more tightly bound perturbations. This results in higher frequency oscillations for perturbing fields around the black hole. Furthermore, the increase in the damping rate with $ Q $ indicates that the perturbations decay more rapidly, meaning that the charged black hole tends to settle down faster after being perturbed. The faster decay is likely due to the increased energy stored in the electromagnetic field, which enhances the dissipation of perturbing waves.

The physical significance of these findings is particularly relevant in the context of astrophysical black holes that may possess charge. Although most observed black holes are expected to have negligible charge due to charge-neutralizing effects in astrophysical environments, highly charged black holes are theoretically possible in certain exotic scenarios. For example, primordial black holes formed in the early universe could retain some charge if they formed in environments where neutralizing particles were absent. Understanding the impact of charge on the QNM spectrum is therefore important for detecting or constraining the properties of such exotic black holes through gravitational wave observations.

These results highlight that the CSBH model exhibits behavior distinct from other well-known charged black hole solutions, such as the Reissner-Nordström black hole. In addition, the impact of the symmergent parameters on the QNM spectrum is distinct from that of the black hole’s charge, further emphasizing the uniqueness of this model. The study of QNMs in charged black holes, and their dependence on parameters such as $ Q $, continues to be an important area of research for both theoretical and observational astrophysics \cite{Yang:2022ifo, Daghigh:2008jz, Zhidenko:2003wq, Zhidenko:2005mv}.

These findings also differ from results obtained for wormhole configurations, as studied in \cite{Ovgun188, gogoi_wormhole}, demonstrating that the symmergent model's predictions for QNMs are distinct even when compared to other exotic spacetime geometries. This underscores the importance of QNMs as a tool for probing the underlying nature of black holes and the influence of additional parameters such as charge and symmergent contributions on their dynamical properties.

Moreover, we have plotted some contours in Fig. \ref{QNMs07} to view the variations of QNMs with respect to the model parameters. The contour plots reveal the impact of the parameters $Q$, $\alpha$ and $c_{\rm O}$ on the QNMs of black holes, focusing on both scalar and electromagnetic perturbations. Physically, as mentioned earlier, the real part $\omega_R$ represents the oscillation frequency, while the imaginary part $\omega_I$ signifies the damping rate, with more negative $\omega_I$ implying faster decay of the perturbations. As $\alpha$ increases, both scalar and electromagnetic oscillation frequencies rise, and the damping rate also increases, meaning perturbations persist shorter for larger $\alpha$. Similarly, in the case of both perturbations, increasing $c_{\rm O}$ from negative to zero and zero to positive values leads to higher oscillation frequencies and the perturbations decay more rapidly. 

The contours displayed in Fig. \ref{QNMs07} also represent the relationship between the real parts of QNMs and the charge $ Q $ for scalar and electromagnetic perturbations, respectively. These contours are crucial for understanding how the charge of a black hole affects the oscillatory frequencies of perturbations. 

The real QNMs signify the oscillation frequencies of scalar field perturbations. As $ Q $ increases, the oscillatory frequency typically increases, reflecting how the black hole’s charge influences the dynamics of scalar perturbations in its spacetime. This result agrees well with our previous depictions in the time domain profiles.
On the other hand, in the behavior of electromagnetic perturbations under similar conditions, the real QNM frequencies here reveal how the black hole’s charge impacts the propagation of electromagnetic waves. Both contours highlight that larger charges lead to higher oscillatory frequencies, demonstrating the charge's stabilizing role in the black hole perturbation dynamics.
These results provide key insights into the stability and resonances of charged black holes, showing that the nature of the perturbing field (scalar vs. electromagnetic) alters the effect of charge on the quasinormal modes, a fundamental aspect in black hole physics and wave propagation in curved spacetime.

These results highlight how the parameters $Q$, $\alpha$ and $c_{\rm O}$ control the stability and longevity of perturbations, which is crucial for understanding the dynamics of black holes and their gravitational wave signatures.

\section{Greybody factors }

The concept of greybody factors originates from Hawking's groundbreaking discovery in 1975, which demonstrated that black holes are not completely black but emit radiation, now known as Hawking radiation \cite{Hawking:1975vcx}. This emission occurs near the black hole's horizon, but the radiation that reaches a distant observer is altered by a redshift factor, resulting in a modification of its spectrum. This distortion is known as the greybody factor, and it reflects the difference between the initial Hawking radiation and what an observer at infinity detects \cite{Singleton:2011vh, Akhmedova:2008dz}. Various methods have been developed to calculate greybody factors, including significant contributions by Maldacena et al. \cite{Maldacena:1996ix}, Fernando \cite{Fernando:2004ay}, and others \cite{Okyay2022, Ovgun188, Pantig:2022gih, Yang:2022xxh, Yang:2022ifo, Panotopoulos:2018pvu, Panotopoulos:2016wuu, Rincon:2018ktz, Ahmed:2016lou, Javed:2022kzf, Al-Badawi:2022aby}.

\subsection{Using the WKB approach}
In subsection, we utilize the higher-order WKB approximation method to calculate greybody factors for both scalar and electromagnetic perturbations. This method is especially suited to calculating the reflection and transmission coefficients associated with wave scattering near a black hole. Specifically, we explore the wave equation under boundary conditions that allow for incoming waves from infinity, a scenario analogous to scattering waves from the black hole's horizon. This allows us to determine how much of the incident wave is reflected back versus transmitted across the potential barrier. The transmission coefficient, in this context, is defined as the greybody factor.

The boundary conditions governing the scattering process are expressed as:
\begin{equation}
\Psi = e^{-i\omega r_*} + R e^{i\omega r_*} \quad \text{as} \quad r_* \rightarrow +\infty, \quad \Psi = T e^{-i\omega r_*} \quad \text{as} \quad r_* \rightarrow -\infty,    
\end{equation}
where $ R $ and $ T $ represent the reflection and transmission coefficients, respectively. These coefficients satisfy the conservation relation $ |T|^2 + |R|^2 = 1 $, implying that the reflection and transmission probabilities must sum to one. The transmission coefficient $ |T|^2 $, also known as the greybody factor $ A $, quantifies the fraction of radiation that escapes the potential barrier and reaches an observer at infinity:
\begin{equation}
|A|^2 = 1 - |R|^2 = |T|^2.    
\end{equation}

The WKB approximation provides a means to calculate the reflection coefficient $ R $ through the following expression:
\begin{equation}
R = \left( 1 + e^{-2 i \pi K} \right)^{-\frac{1}{2}},    
\end{equation}
where the phase factor $ K $ is determined by the following equation:
\begin{equation}
K - i \frac{(\omega^2 - V_{0})}{\sqrt{-2 V_{0}^{\prime \prime}}} - \sum_{i=2}^{i=6} \Lambda_{i}(K) =0.
\end{equation}
which involves the maximum value of the effective potential $ V_0 $, its second derivative $ V_{0}^{\prime \prime} $, and higher-order corrections $ \Lambda_i $, extending up to the 6th order \cite{Schutz, Will_wkb, Konoplya_wkb, Maty_wkb}. The WKB method, while effective, is less accurate at low frequencies where reflection tends to be total, causing the greybody factors to approach zero. Nevertheless, this does not substantially affect the calculation of energy emission rates.

Given the robustness and broad applicability of the WKB method in various contexts, including the study of black hole perturbations and greybody factors, a detailed review is beyond the scope of this paper. For more in-depth discussions, we refer readers to comprehensive reviews in the literature \cite{Konoplya:2019hlu, Konoplya:2011qq}.

\begin{figure}[htbp]
\centerline{
   \includegraphics[scale = 0.8]{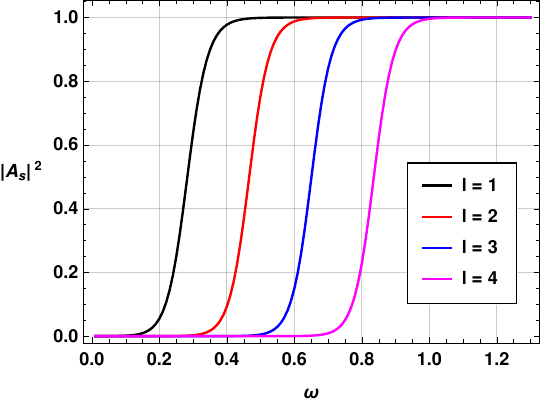}\hspace{0.5cm}
   \includegraphics[scale = 0.8]{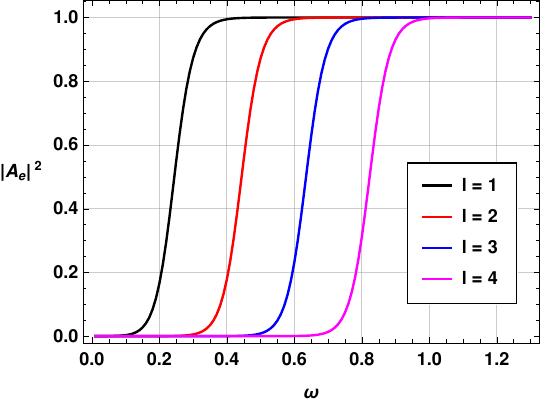}} \vspace{-0.2cm}
\caption{The greybody factors for massless scalar (left panel) and electromagnetic (right panel) perturbations for different values of the multipole moment $l$ with the parameter values $M=1$, $G = 1$, $\alpha = 0.9$, $Q = 0.3$ and $c_{\rm O} = 0.4$.}

\label{G01}
\end{figure}

\begin{figure}[htbp]
\centerline{
   \includegraphics[scale = 0.8]{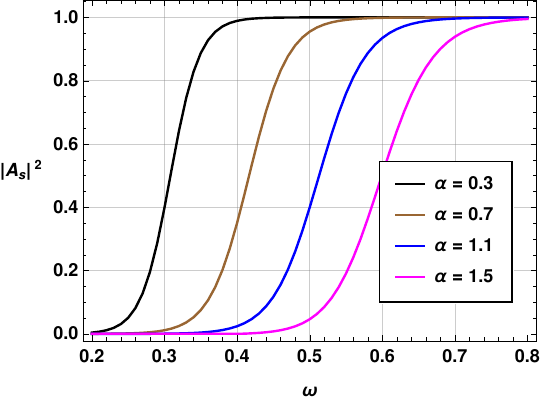}\hspace{0.5cm}
   \includegraphics[scale = 0.8]{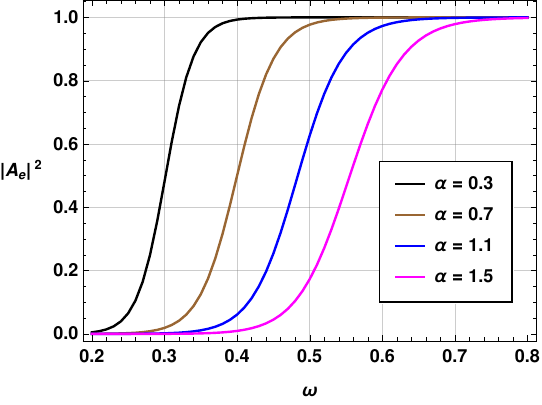}} \vspace{-0.2cm}
\caption{The greybody factors for massless scalar (left panel) and electromagnetic (right panel) perturbations for different values of the vacuum energy parameter $\alpha$ with the parameter values $M=1$, $G = 1$, $l = 2$, $Q = 0.3$ and $c_{\rm O} = 0.1$.}

\label{G02}
\end{figure}

\begin{figure}[htbp]
\centerline{
   \includegraphics[scale = 0.8]{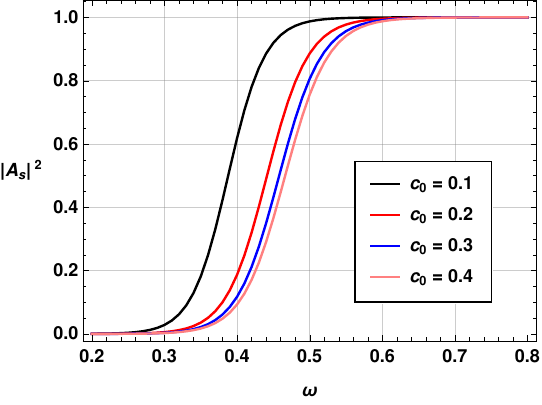}\hspace{0.5cm}
   \includegraphics[scale = 0.8]{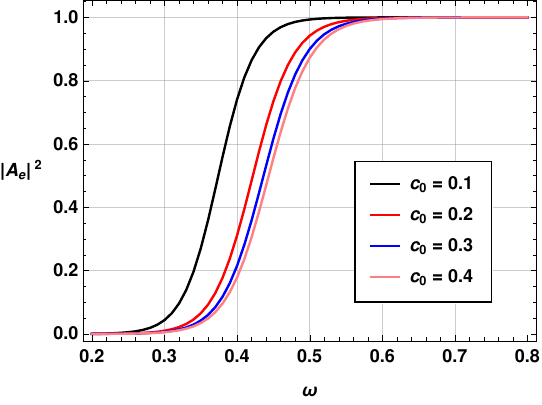}} \vspace{-0.2cm}
\caption{The greybody factors for massless scalar (left panel) and electromagnetic (right panel) perturbations for different values of the symmergent parameter $c_{\rm O}$ with the parameter values $M=1$, $G = 1$, $\alpha = 0.9, Q = 0.3$ and $l = 2$.}

\label{G03}
\end{figure}

The figures presented in \ref{G01}, \ref{G02}, \ref{G03} and \ref{G04} provide a detailed analysis of the behavior of the greybody factors for massless scalar and electromagnetic perturbations under varying model parameters. Each figure highlights how different parameters influence the absorption probabilities, offering valuable insights into black hole dynamics in the context of symmergent gravity.

In Fig. \ref{G01}, we observe the variation of greybody factors $ |A_s|^2 $ for scalar perturbations (left panel) and $ |A_e|^2 $ for electromagnetic perturbations (right panel) as the multipole moment $ l $ changes. As the multipole moment increases, the peak of the greybody factors shifts to higher frequencies ($ \omega $), indicating that higher-energy modes are more strongly excited at larger $ l $ values. Interestingly, electromagnetic perturbations reach their maximum absorption at slightly lower frequencies compared to scalar perturbations. This suggests that electromagnetic waves interact more efficiently at lower frequencies, leading to earlier peaks in absorption than scalar waves for the same multipole moments.

In Fig. \ref{G02}, the effects of the model parameter $ \alpha $ on the greybody factors are examined. The greybody factors for both scalar and electromagnetic perturbations exhibit a noticeable decrease as $ \alpha $ increases. This implies that black holes with smaller $ \alpha $ values have higher absorption and scattering probabilities, effectively capturing more incoming radiation or particles. As $ \alpha $ increases, the greybody factors drop, indicating that black holes become less interactive with the surrounding radiation and allow more of it to escape. The sensitivity of the greybody factors to $ \alpha $ underscores the significance of this parameter in controlling the interaction between black holes and external perturbations.

A similar pattern emerges in Fig. \ref{G03}, which explores the impact of the model parameter $c_{\rm O}$ on the greybody factors. Both scalar and electromagnetic perturbations display higher greybody factors for smaller values of $c_{\rm O}$, with a clear decrease as $c_{\rm O}$ increases. This suggests that black holes with lower values of $c_{\rm O}$ are more effective at absorbing radiation, while higher values of $c_{\rm O}$ reduce this efficiency. The sensitivity of the greybody factors to $c_{\rm O}$ diminishes at higher values, indicating that changes in $c_{\rm O}$ have less impact on black hole absorption in this regime.

Finally, Fig. \ref{G04} illustrates the influence of the charge parameter $ Q $ on the greybody factors. As the charge $ Q $ increases, the greybody factors gradually decrease for both types of perturbations. This implies that charged black holes are less efficient at absorbing radiation compared to their neutral counterparts. The decrease in absorption with increasing charge suggests that the presence of charge reduces the interaction between the black hole and incoming waves or particles, leading to lower absorption probabilities.

In summary, these figures reveal that the parameters $ l $, $ \alpha $, $c_{\rm O}$, and $ Q $ significantly influence the greybody factors for both scalar and electromagnetic perturbations. Smaller values of $ \alpha $, $c_{\rm O}$, and $ Q $ correspond to higher absorption probabilities, while higher values lead to decreased interaction with incoming radiation. Additionally, electromagnetic perturbations tend to reach their maximum absorption at lower frequencies compared to scalar perturbations. These findings enhance our understanding of the intricate dynamics between black holes and external perturbations within the framework of symmergent gravity.

\begin{figure}[htbp]
\centerline{
   \includegraphics[scale = 0.8]{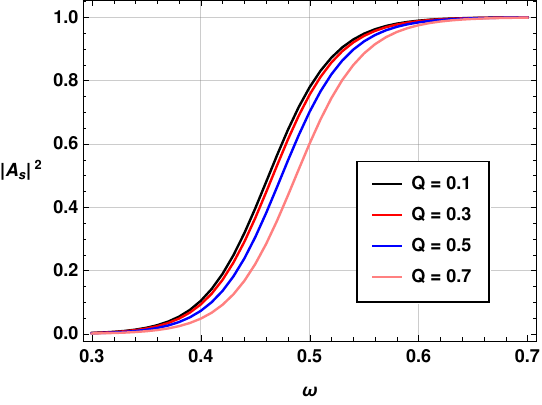}\hspace{0.5cm}
   \includegraphics[scale = 0.8]{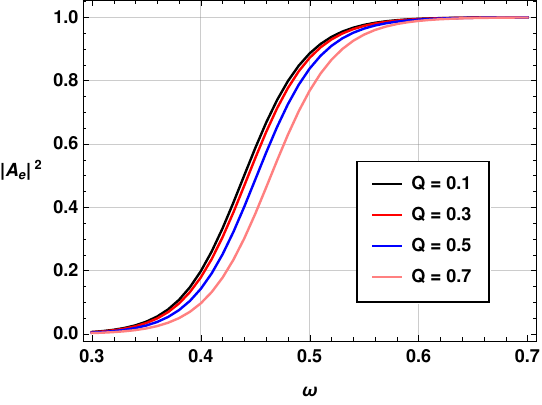}} \vspace{-0.2cm}
\caption{The greybody factors for massless scalar (left panel) and electromagnetic (right panel) perturbations for different values of the charge parameter $Q$ with the parameter values $M=1$, $G = 1$, $\alpha = 0.9, c_{\rm O} = 0.4$ and $l = 2$.}
\label{G04}
\end{figure}

\subsection{Rigorous Bounds on Greybody Factors} \label{sec5}

In this part of our investigation, we consider rigorous bounds on greybody factors by utilizing a different method. Since this portion of the analysis reveals similar behavior between scalar and electromagnetic perturbations in terms of greybody factors, we focus on scalar perturbations only for the remainder of the study. 

The method we employ is based on the elegant analytical approach originally introduced by Visser (1998) \cite{Visser:1998ke}, which was subsequently developed further by Boonserm and Visser (2008) \cite{Boonserm:2008zg}. This technique has been widely applied in various contexts, as explored by numerous researchers, including Boonserm et al. (2017, 2019) \cite{Boonserm:2017qcq}, Yang et al. (2022) \cite{Yang:2022ifo}, Gray et al. (2015) \cite{Gray:2015xig}, Ngampitipan et al. (2012) \cite{Ngampitipan:2012dq}, and others \cite{Chowdhury:2020bdi,Miao:2017jtr,Liu:2021xfs,Barman:2019vst}. These studies have demonstrated the robustness of this approach in determining greybody factor bounds across different gravitational systems.

For our analysis, we concentrate specifically on deriving the bounds for the greybody factors of CSBHs. We begin by analyzing the Klein-Gordon equation for the massless scalar field, as discussed in the previous sections, and then reduce the effective potential to the form:
\begin{equation}
    V(r) = \frac{l(l + 1)h(r)}{r^{2}} + \frac{h(r)h'(r)}{r},
\end{equation}
where $ h(r) $ represents the metric function and $ l $ is the multipole moment. This potential governs the dynamics of scalar field perturbations in the black hole background.

Using this effective potential, we proceed to derive the rigorous lower bound for the greybody factors, following the methodology laid out by Visser (1998) \cite{Visser:1998ke} and Boonserm and Visser (2008) \cite{Boonserm:2008zg}. The bound for the transmission coefficient, denoted as $ T_b $, is given by:
\begin{equation}
    T_b \geq \operatorname{sech}^{2}\left(\frac{1}{2 \omega} \int_{-\infty}^{\infty}\left|V\right| \frac{d r}{h(r)} \right),
\end{equation}
where $ \omega $ is the frequency of the perturbation. Here, $ T_b $ represents the transmission coefficient, which corresponds to the greybody factor.

To account for the presence of the cosmological constant and the symmergent gravity parameters, we modify the boundary conditions in accordance with the work by Boonserm et al. (2019) \cite{Boonserm:2019mon}. The modified bound is expressed as:
\begin{equation}
    A \geq T_b=\operatorname{sech}^{2}\left(\frac{1}{2 \omega} \int_{r_{H}}^{R_{H}} \frac{|V|}{h(r)} d r\right)=\operatorname{sech}^{2}\left(\frac{A_{l}}{2 \omega}\right),
\end{equation}
where the integral term $ A_{l} $ is defined as:
\begin{equation}
    A_{l}=\int_{r_{H}}^{R_{H}} \frac{|V|}{h(r)} d r=\int_{r_{H}}^{R_{H}}\left|\frac{l(l+1)}{r^{2}}+\frac{f^{\prime}}{r}\right| d r.
\end{equation}
In these equations, $ r_H $ and $ R_H $ denote the event and cosmological horizon radii of the black hole, and the effective potential $ V(r) $ is integrated between these radii.

We have successfully computed the rigorous bounds on greybody factors for CSBHs. The formula obtained is expressed as:
\begin{equation}
T_b=\text{sech}^2\left(\frac{\left(R_H-r_H\right) \left(4 \pi  c_{\rm O} G \left(3 \alpha  r_H R_H \left(G M \left(r_H+R_H\right)+l (l+1) r_H R_H\right)-Q^2 C_1\right)+(\alpha -1) \alpha  r_H^3 R_H^3\right)}{24 \omega \pi  \alpha  c_{\rm O} G r_H^3 R_H^3}\right)
\end{equation}
where $C_1 = \left(r_H R_H+r_H^2+R_H^2\right)$ and the parameters $ c_{\rm O} $, $ \alpha $, and $ Q $ represent the quadratic curvature coupling parameter, the vacuum energy parameter, and the charge of the black hole, respectively.

This expression provides a rigorous lower bound on the greybody factors as a function of various parameters. Our findings demonstrate that the greybody factor bound is sensitive to both the black hole charge and the parameters governing the symmergent gravity model. These results will contribute to a deeper understanding of black hole radiation and the role of symmergent gravity in shaping the interaction between black holes and perturbations.

\begin{figure}[htbp]
\centerline{
   \includegraphics[scale = 0.8]{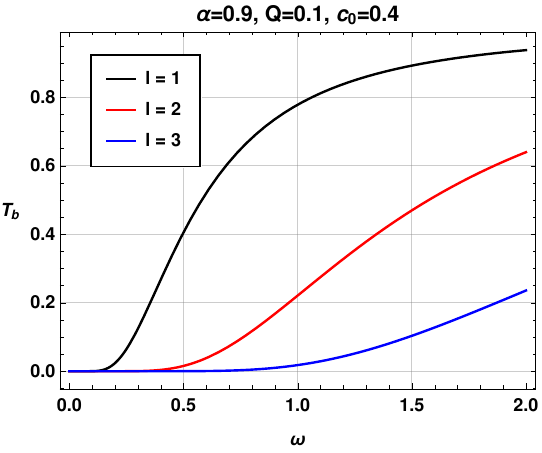}\hspace{0.5cm}
   \includegraphics[scale = 0.8]{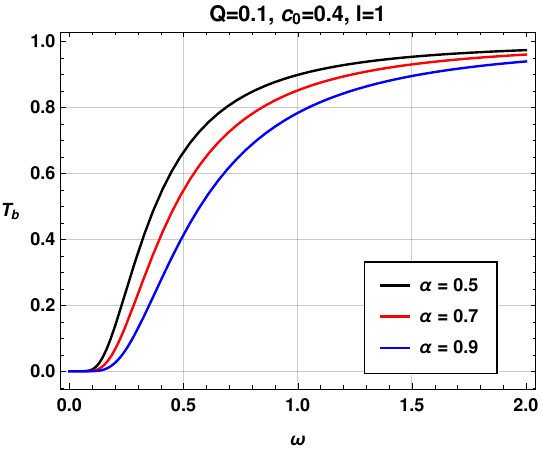}} \vspace{-0.2cm}
   
\caption{The greybody bound $T_b$ as a function of the frequency.}
\label{G04}
\end{figure}

\begin{figure}[htbp]
\centerline{
   \includegraphics[scale = 0.8]{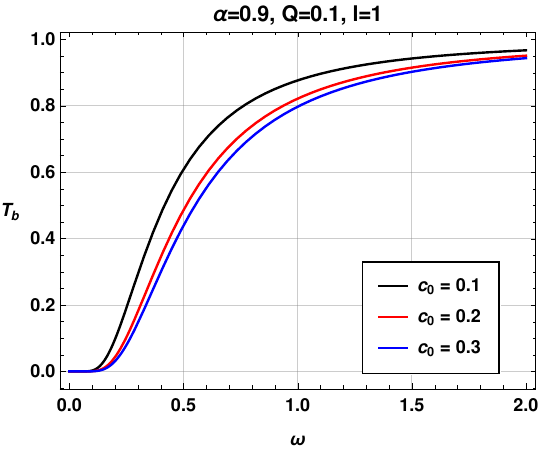}\hspace{0.5cm}
   \includegraphics[scale = 0.8]{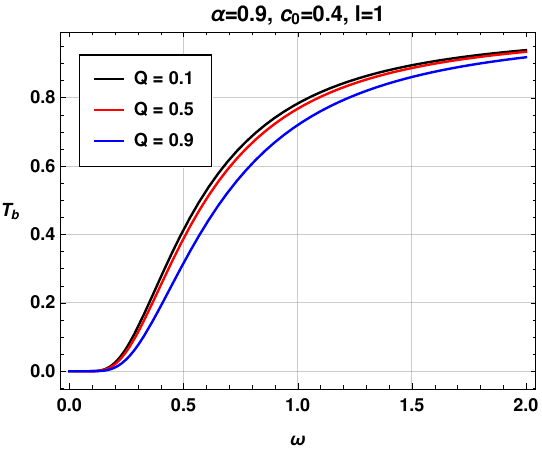}} \vspace{-0.2cm}
   
\caption{The greybody bound $T_b$ as a function of the frequency.}
\label{G05}
\end{figure}

By performing numerical calculations, we can evaluate the bound and visualize it in Fig. \ref{G04} for the case of different $l$ values (on the left panel) and different $\alpha$ values (on the right panel), and in Fig. \ref{G05} for the case of different $c_{\rm O}$ values (on the left panel) and different $Q$ values (on the right panel). The resulting graphs indicate that with an increase in the model parameter $\alpha$, greybody bounds decrease significantly. The impacts of parameter $l$ on the greybody bounds are more significant in comparison to the greybody factors discussed in the previous subsection. Again, as the parameter $c_{\rm O}$ increases, the bound on the greybody factor decreases. This observation suggests that CSBHs exhibit stronger barrier properties and possess lower greybody bounds compared to Schwarzschild black holes. The model parameter $Q$ also has a similar impact on the greybody bounds. Overall, the influence of the model parameters for CSBHs on the greybody bounds almost mirrors the effects observed in the case of greybody factors discussed in the previous subsection. Given that greybody factors exhibit similar behavior for both types of perturbations, in this subsection, we have focused solely on scalar perturbations to investigate the greybody bounds.

\section{Concluding Remarks}

In this paper, we have studied charged black hole solutions in the Symmergent gravity \cite{Pulice:2023dqw} by investigating the QNMs and greybody factors of the CSBHs, focusing on scalar and electromagnetic perturbations. This work is an extension of our previous work \cite{bh5}. Here, we have analysed the QNMs and greybody factors in the presence of charge $Q$. Our analyses reveal that the symmergent parameters $\alpha$, and the quadratic curvature coupling parameter $c_{\rm O}$ have significant impacts on the QNMs and greybody factors as well as the black hole charge $Q$ and the multipole moment $l$.
\\ \\
\textbf{Key insights from our results include:}
\begin{itemize}

	\item	Higher multipole moments $l$ lead to stronger potential barriers, higher QNM frequencies, and faster damping rates, with black holes becoming more resistant to perturbations.
\item	The symmergent parameter $\alpha$ shows a linear increase in both the QNM frequency and damping rate. This suggests that the parameter governs the black hole’s ability to trap and damp perturbations, particularly in scalar perturbations.
	\item The coupling parameter $c_{\rm O}$ also plays a critical role, exhibiting a complex influence on the QNM spectrum and the greybody factors. At small values, rapid changes occur in both oscillation frequency and damping rates, but these effects stabilize as $c_{\rm O}$ increases.
	\item The black hole charge $Q$ increases the potential barrier, resulting in higher frequencies for QNMs, while marginally affecting the damping rate.
 \end{itemize}

The greybody factors of CSBHs were found to differ substantially from those of Schwarzschild black holes. Particularly, the model parameters $\alpha$ and $c_{\rm O}$ play a central role in determining the black hole’s absorption and scattering properties. We observed that smaller values of these parameters correspond to higher greybody factors, indicating stronger interactions between the black hole and incoming matter or waves.

Our analysis shows that small positive values of  $c_{\rm O}$  result in smaller real QNM frequencies, while small negative values lead to larger frequencies. For asymptotically large positive or negative values of  $c_{\rm O}$, the QNMs approach constant values similar to those in a Schwarzschild black hole. The parameter  $\alpha$  has a near-linear effect on both the real and imaginary components of the QNMs, with larger values of $\alpha$ leading to higher frequencies and faster decay rates. Additionally, we observe that the presence of charge  $Q$  significantly impacts the QNMs by raising both the frequency and damping rate as $Q$ increases.

Our investigation into scalar and vector perturbations shows that scalar perturbations consistently exhibit higher frequencies and shorter lifetimes compared to electromagnetic perturbations, a reflection of the stronger interaction between the scalar field and the black hole. The time-domain profiles further highlight that scalar perturbations decay more rapidly, with notable sensitivity to variations in  $c_{\rm O}$,  $\alpha$, and $Q$ .

Regarding the greybody factors, our study indicates that both  $\alpha$  and  $c_{\rm O}$  have measurable effects on the absorption and scattering behavior of the black hole. We find that smaller values of  $\alpha$  increase the greybody factors, suggesting that black holes with reduced  $\alpha$  are more effective in absorbing and scattering incoming radiation.

These results provide deeper insight into the role of symmergent gravity parameters in shaping the dynamical and observational properties of charged black holes. The distinct signatures of these parameters, particularly in QNM spectra and greybody factors, offer potential observational targets for future gravitational wave and black hole shadow experiments. With the advent of sensitive detectors like LISA and ongoing observations from the Event Horizon Telescope (EHT), it may soon be possible to use these findings to place further constraints on the symmergent gravity and improve our understanding of black hole physics. Our present work, along with the observational results of QNMs can help us to understand symmergent gravity in more detail. Observational constraints on QNMs in the symmergent gravity, derived from LISA data, alongside shadow constraints from the EHT, could soon provide a means to test the consistency and viability of the theory.

\acknowledgements
 A. {\"O}., D. J. G. and B. P. would like to acknowledge networking support by the COST Action CA21106 - COSMIC WISPers in the Dark Universe: Theory, astrophysics and experiments (CosmicWISPers). The work of B. P. is supported by Sabanc{\i} University, Faculty of Engineering and Natural Sciences. The work of B. P. is supported by the Astrophysics Research Center of the Open University of Israel (ARCO) through The Israeli Ministry of Regional Cooperation. 

%%%%%%%%%%%%%%%%%%%%%%%%%%%
%%  PLEASE KEEP EACH BIBITEM IN A SINGLE LINE WITHOUT USING 'ENTER'
%%%%%%%%%%%%%%%%%%%%%%%%%%%

%NO NEED DOIS.

\end{document}